\documentclass[journal]{IEEEtran}
\usepackage{amsmath,amsfonts}
\usepackage{algorithmic}
\usepackage{algorithm}
\usepackage{array}
\usepackage[caption=false,font=footnotesize,labelfont=rm,textfont=rm]{subfig}
\usepackage{textcomp}
\usepackage{stfloats}
\usepackage{url}
\usepackage{verbatim}
\usepackage{graphicx}
\usepackage{cite}
\usepackage{amsthm}
\usepackage{amssymb}
\usepackage{epsfig}
\usepackage{graphicx}
\usepackage{balance}
\usepackage{epstopdf}
\usepackage{multicol}
\usepackage{multirow}
\usepackage{threeparttable}
\usepackage{booktabs}
\usepackage{cite}
\usepackage{cases}
\usepackage{cuted}
\usepackage{xcolor}
\usepackage{threeparttable}
\usepackage{amsmath}
\allowdisplaybreaks[2]
\usepackage{float}
\usepackage{placeins}

\begin{document}

\title{Dynamic Multiple-Parameter Joint Time-Vertex Fractional Fourier Transform and its Intelligent Filtering Methods}

\author{Manjun~Cui, Ziqi~Yan, Yangfan~He, and Zhichao~Zhang,~\IEEEmembership{Member,~IEEE}
	\thanks{This work was supported in part by the Open Foundation of Hubei Key Laboratory of Applied Mathematics (Hubei University) under Grant HBAM202404; in part by the Foundation of Key Laboratory of System Control and Information Processing, Ministry of Education under Grant Scip20240121; and in part by the Startup Foundation for Introducing Talent of Nanjing Institute of Technology under Grant YKJ202214. \emph{(Corresponding author: Zhichao~Zhang.)}}
	\thanks{Manjun~Cui and Ziqi~Yan are with the School of Mathematics and Statistics, Nanjing University of Information Science and Technology, Nanjing 210044, China (e-mail: cmj1109@163.com; yanziqi54@gmail.com).}
	\thanks{Yangfan~He is with the School of Communication and Artificial Intelligence, School of Integrated Circuits, Nanjing Institute of Technology, Nanjing 211167, China (e-mail: Yangfan.He@njit.edu.cn).}
	\thanks{Zhichao~Zhang is with the School of Mathematics and Statistics, Nanjing University of Information Science and Technology, Nanjing 210044, China, with the Hubei Key Laboratory of Applied Mathematics, Hubei University, Wuhan 430062, China, and also with the Key Laboratory of System Control and Information Processing, Ministry of Education, Shanghai Jiao Tong University, Shanghai 200240, China (e-mail: zzc910731@163.com).}}

\markboth{IEEE TRANSACTIONS ON PATTERN ANALYSIS AND MACHINE INTELLIGENCE}
{Shell \MakeLowercase{\textit{et al.}}: Bare Demo of IEEEtran.cls for Journals}


\maketitle

\begin{abstract}
Dynamic graph signal processing provides a principled framework for analyzing time-varying data defined on irregular graph domains. However, existing joint time-vertex transforms such as the joint time-vertex fractional Fourier transform assign only one fractional order to the spatial domain and another one to the temporal domain, thereby restricting their capacity to model the complex and continuously varying dynamics of graph signals. To address this limitation, we propose a novel dynamic multiple-parameter joint time-vertex fractional Fourier transform (DMPJFRFT) framework, which introduces time-varying fractional parameters to achieve adaptive spectral modeling of dynamic graph structures. By assigning distinct fractional orders to each time step, the proposed transform enables dynamic and flexible representation of spatio-temporal signal evolution in the joint time-vertex spectral domain. Theoretical properties of the DMPJFRFT are systematically analyzed, and two filtering approaches: a gradient descent-based method and a neural network-based method, are developed for dynamic signal restoration. Experimental results on dynamic graph and video datasets demonstrate that the proposed framework effectively captures temporal topology variations and achieves superior performance in denoising and deblurring tasks compared with some state-of-the-art graph-based transforms and neural networks.
\end{abstract}


\section{Introduction}
\indent As data becomes increasingly structured in complex systems such as social networks, transportation systems, sensor networks, and bioinformatics networks, traditional signal processing (SP) methods defined in Euclidean space are no longer able to effectively characterize the irregular relationships inherent in such data. Graph signal processing (GSP) has emerged as a systematic theoretical framework for defining, analyzing, and processing signals on graphs \cite{ortega22introduction,Leus23,Ortega08}. In recent years, GSP has demonstrated broad application potential in fields such as social relationship analysis, recommender systems, traffic flow prediction, and brain functional connectivity modeling \cite{Saboksayr21,Mateos19Connecting,Colonnese21Protein,Ma16Diffusion,Xu24Dynamic,Qu22Brain}.

\indent Within the framework of GSP, spectral analysis plays a crucial role in characterizing the structural and functional properties of graph signals. Among the commonly used spectral methods, the graph Fourier transform (GFT) \cite{Sandryhaila13,Gavili17,Lu19,Domingos20,Patane23,Qi22} and the graph fractional Fourier transform (GFRFT) \cite{Wang18,Ozturk21,yan2021windowed,gan2025windowed,Alik24Wiener,alikacsifouglu2025joint} enable the extraction of spatial-frequency features from graph signals. However, these methods are limited in that they cannot directly capture temporal-frequency characteristics, which are essential for modeling signals that evolve over time. To jointly extract spatio-temporal frequency characteristics, researchers proposed the joint time-vertex Fourier transform (JFT) \cite{Grassi18JFT,Loukas16Frequency}. Later, the fractional-order concept is introduced to extend JFT into the joint time-vertex fractional Fourier transform (JFRFT) \cite{Alik24Wiener,alikacsifouglu2025joint,yan2025trainable}, which provides a parameterized spectral transformation. Despite these improvements, existing JFRFT typically employs a single fractional order for the spatial domain and another single fractional order for the temporal domain. This restricts their ability to adaptively model heterogeneous features across different vertex clusters, temporal segments, or frequency bands.

\indent Inspired by the concepts of the multiple-parameter GFRFT (MPGFRFT) \cite{cui2025multiple} and multiple-parameter discrete FRFT (MPDFRFT) \cite{kang16multiple}, it is natural to consider introducing multiple fractional orders to control different spectral components, thereby enhancing the expressiveness and adaptability of the transform. However, directly replacing the GFRFT operator in the JFRFT with an MPGFRFT operator is insufficient. Although this modification introduces parameter diversity in the graph spectral domain, it still employs the same set of parameters across the entire temporal dimension, thus failing to fully exploit temporal variability. To effectively capture the spatio-temporal evolution characteristics of dynamic graph signals, it is therefore necessary to design a framework of dynamic multiple-parameter JFRFT (DMPJFRFT) that applies different MPGFRFTs to distinct times within the joint time-vertex fractional spectral domain, enabling a more flexible fractional spectral representation that accounts for the intrinsic characteristics of dynamic graphs.

\indent In GSP, filtering serves as one of the most fundamental and essential operations, enabling noise suppression, signal recovery, and feature enhancement through spectral domain manipulation \cite{Ozturk21,Gavili17,Alik24Wiener,Bianchi22Graph,Chen24,Isufi24,Patane23}. However, filtering dynamic graph signals remains challenging due to their complex and time-varying structures. The proposed DMPJFRFT framework provides a new opportunity to perform adaptive filtering in the joint time-vertex fractional spectral domain, where varying fractional orders enable dynamic graph structure to be effectively modeled in the graph spectral representation. This flexibility facilitates improved filtering performance, including enhanced denoising and deblurring of dynamic graph signals.

\indent The main contributions of this paper are summarized as follows:
\begin{itemize}
	\item We propose four variants of the DMPJFRFT and establish their theoretical framework for effectively modeling dynamic graph signals.
	
	\item We develop a gradient descent-based DMPJFRFT filtering method for effective signal filtering, including denoising and deblurring.
	
	\item We introduce a neural network-based DMPJFRFT filtering approach that leverages partial prior information to achieve adaptive filtering with enhanced performance and generalization capabilities.
\end{itemize}

\begin{figure*}[htbp]
	\centering
	\includegraphics[width=5in]{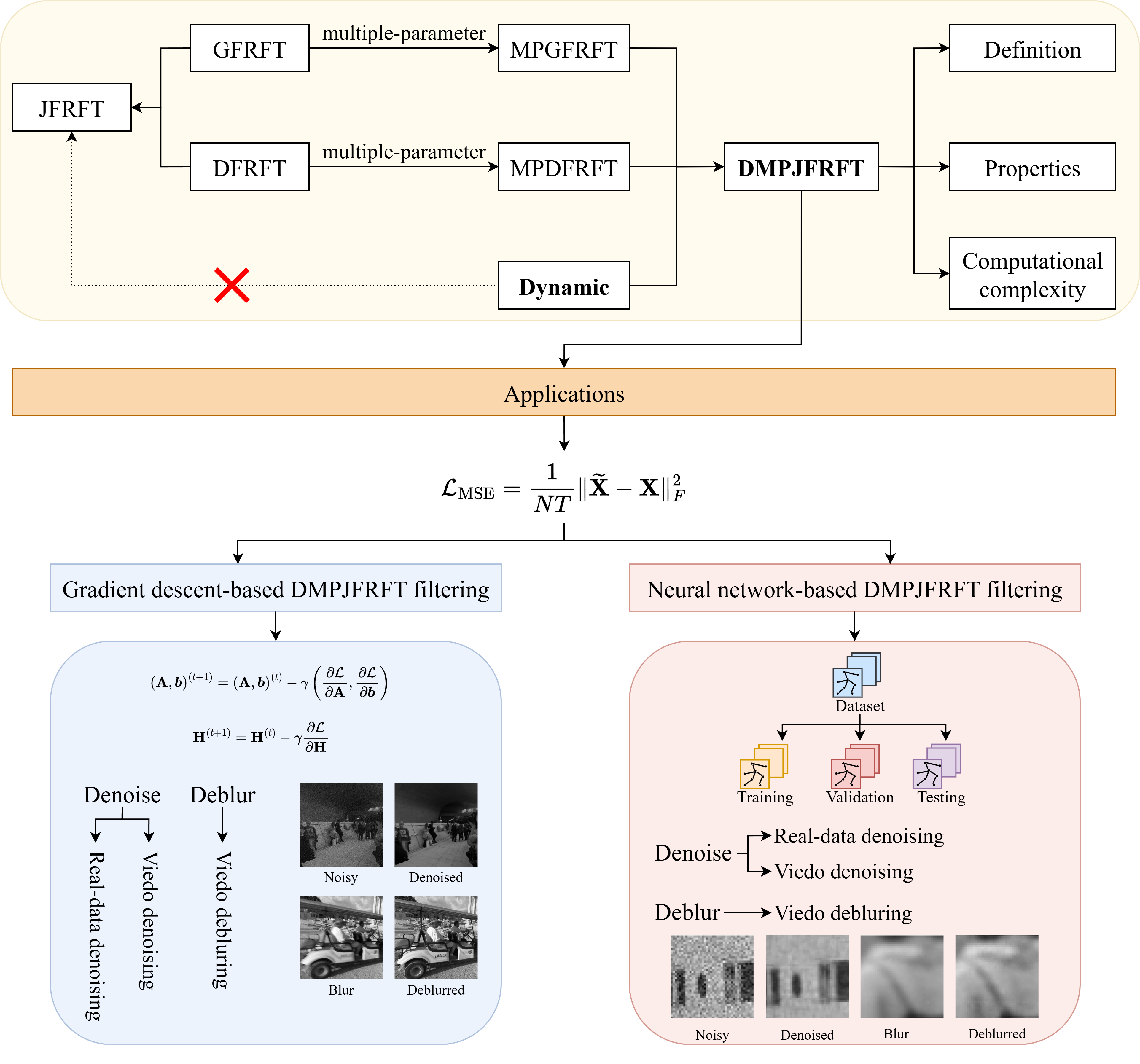}
	\caption{Overall framework of the proposed DMPJFRFT-based dynamic graph signal processing model.}
	\label{fig:DMPJFRFT}
\end{figure*}

\indent The remainder of this paper is organized as follows. Section \ref{sec2} introduces some necessary preliminary knowledge. Section \ref{sec3} presents the four variants of DMPJFRFT along with their theoretical framework. Section \ref{sec4} proposes the gradient descent-based DMPJFRFT filtering method and presents the corresponding simulation results. Section \ref{sec5} introduces the neural network-based DMPJFRFT filtering approach and presents its simulation results. Section \ref{sec6} concludes the paper. To provide a clear overview of the proposed framework and its key components, the overall structure of this study is illustrated in Fig.~\ref{fig:DMPJFRFT}. All the technical proofs of our theoretical results are relegated to the Appendix parts.

\section{Preliminaries} \label{sec2}
\subsection{Joint Time-Vertex Fractional Fourier Transform}
\indent Let $\mathcal{G}=\left\{\mathcal{N},\mathcal{E},\mathbf{A}\right\}$ denote a graph with vertex set $\mathcal{N}$ of cardinality $|\mathcal{N}| = N$, and adjacency matrix $\mathbf{A} \in \mathbb{R}^{N \times N}$. A static graph signal is represented by $\mathbf{x}=[x_0,x_1,\cdots,x_{N-1}]^{\mathrm{T}} \in \mathbb{C}^{N}$, where $x_i$ denotes the signal value at the $i$-th node . 

\indent Let $\mathbf{Z}$ be a graph shift operator (GSO) (e.g., adjacency, Laplacian, normalized Laplacian, row normalized adjacency, or symmetric normalized adjacency) with Jordan decomposition $\mathbf{Z}=\mathbf{U}_Z \mathbf{J}_Z \mathbf{U}_Z^{-1}$. The GFT matrix is defined as $\mathbf{F}_G=\mathbf{U}_Z^{-1}$, which admits the spectral decomposition $\mathbf{F}_G=\mathbf{V}_G\mathbf{\Lambda}_G\mathbf{V}_G^{-1}$. The GFRFT matrix of order $\alpha \in \mathbb{R}$ is then defined as
\begin{equation}
	\mathbf{F}_G^{\alpha} = \mathbf{V}_G\mathbf{\Lambda}_G^{\alpha} \mathbf{V}_G^{-1}.
\end{equation}
This matrix reduces to the identity matrix $\mathbf{I}_N$ for $\alpha = 0$ and to the GFT matrix $\mathbf{F}_G$ for $\alpha = 1$.

\indent The DFRFT matirx of order $\beta$ is
\begin{equation}
	\mathbf{D}_T^{\beta} = \mathbf{V}_T \mathbf{\Lambda}_T^{\beta} \mathbf{V}_T^{-1},
\end{equation}
where $\mathbf{V}_T$ and $\mathbf{\Lambda}_T$ are the eigenvectors and eigenvalues of the DFT matrix $\mathbf{D}_T$, respectively. Likewise, $\beta=0$ yields the identity matrix $\mathbf{I}_N$ and $\beta=1$ recovers the DFT matrix $\mathbf{D}_T$.

\indent For a time-varying graph signal $\mathbf{X} = [\mathbf{x}_1, \dots, \mathbf{x}_T] \in \mathbb{C}^{N \times T}$, where $N$ and $T$ denote the numbers of vertices and time instances, respectively, the JFRFT of $\mathbf{X}$ with order pair $(\alpha,\beta) \in \mathbb{R}^2$ is defined as\cite{Alik24Wiener,alikacsifouglu2025joint}
\begin{equation}
	\mathrm{JFRFT}_G^{(\alpha,\beta)}(\mathbf{X})=\mathbf{F}_G^{\alpha}\mathbf{X}\left(\mathbf{D}_T^{\beta}\right)^{\mathrm{T}}.
\end{equation}
The JFRFT provides a parametric transformation between the time-vertex domain and the joint spectral domain, where the parameters $(\alpha,\beta)$ control the fractional progression between the identity transform and the full joint Fourier transform. When $(\alpha,\beta) = (0,0)$, the transform reduces to the identity operator, whereas $(\alpha,\beta) = (1,1)$ yields the JFT.

\indent For the vectorized signal $\mathbf{x} = \mathrm{vec}(\mathbf{X}) \in \mathbb{C}^{NT}$, the JFRFT can be equivalently expressed in matrix form as
\begin{equation}
	\mathrm{JFRFT}_G^{(\alpha,\beta)}(\mathbf{x})=\mathbf{F}_J^{(\alpha,\beta)} \mathbf{x}=
	\left(\mathbf{D}_T^{\beta} \otimes \mathbf{F}_G^{\alpha}\right) \mathbf{x},
\end{equation}
where $\otimes$ denotes the Kronecker product and $\mathbf{F}_J^{(\alpha,\beta)}$ is JFRFT matrix.

\subsection{Multiple-Parameter Graph Fractional Fourier Transform}
\indent Suppose that $\mathbf{F}_G$ can be orthogonally decomposed that $\mathbf{F}_G = \mathbf{V}_G\boldsymbol{\Lambda}_G\mathbf{V}_G^{-1}$, where $\mathbf{V}_G$ is the matrix of eigenvectors and $\boldsymbol{\Lambda}_G=\mathrm{diag}(\lambda_0,\dots,\lambda_{N-1})$ is the corresponding eigenvalue matrix. For an order vector $\boldsymbol{a} = [a_0,a_1,\dots,a_{N-1}]^{\mathrm{T}} \in \mathbb{R}^{N}$, the multiple-parameter GFRFT (MPGFRFT) allows each graph spectral component to be modulated by an individual fractional order parameter. Two alternative forms of MPGFRFT are defined in \cite{cui2025multiple}.

\indent The type-I MPGFRFT (MPGFRFT-I) of a graph signal $\mathbf{x}\in\mathbb{C}^{N}$ is defined as 
\begin{equation}
	\widehat{\mathbf{x}}^{\mathrm{I}}_{\boldsymbol{a}}=\mathbf{F}_{\mathrm{I}}^{\boldsymbol{a}}\mathbf{x},
\end{equation}
where
\begin{equation}
	\mathbf{F}_{\mathrm{I}}^{\boldsymbol{a}}=\mathbf{V}_G\mathrm{diag}\left(\lambda_0^{a_0},\dots,\lambda_{N-1}^{a_{N-1}}\right)\mathbf{V}_G^{-1}
\end{equation}
is the MPGFRFT-I matrix.

\indent The type-II MPGFRFT (MPGFRFT-II) of a graph signal $\mathbf{x}\in\mathbb{C}^{N}$ is defined as
\begin{equation}
	\widehat{\mathbf{x}}^{\mathrm{II}}_{\boldsymbol{a}}=\mathbf{F}_{\mathrm{II}}^{\boldsymbol{a}}\mathbf{x},
\end{equation}
\begin{equation}
	\mathbf{F}_{\mathrm{II}}^{\boldsymbol{a}}=\sum_{n=0}^{N-1}C_{n,a_n}^{\mathrm{II}}\mathbf{F}_G^{n},
\end{equation}
where $\mathbf{F}_G$ denotes the GFT matrix and the coefficients are
\begin{equation}
	C_{n,a_n}^{\mathrm{II}}
	=
	\sum_{j=0}^{N-1}p_{n+1,j+1}\lambda_j^{a_n},
\end{equation}
with $p_{n+1,j+1}$ being the $(n+1,j+1)$-th entry of the inverse Vandermonde matrix associated with $(\lambda_0,\dots,\lambda_{N-1})$, that is
\begin{align}    \label{eq:MPGFRFT_P}
	\mathbf{P}=\left(p_{i j}\right) \triangleq\left(\begin{array}{cccc}
		1 & \lambda_{0} & \cdots & \lambda_{0}^{N-1} \\
		1 & \lambda_{1} & \cdots & \lambda_{1}^{N-1} \\
		\vdots & \vdots & \ddots & \vdots \\
		1 & \lambda_{N-1} & \cdots & \lambda_{N-1}^{N-1}
	\end{array}\right)^{-1}.
\end{align}

\subsection{Multiple-Parameter Discrete Fractional Fourier Transform} 
\indent For temporal signal $\mathbf{x}=(x_0,x_1,\cdots,x_{T-1})^{\mathrm{T}}\in\mathbb{C}^{T}$, classical spectral analysis is usually carried out via the DFRFT. In~\cite{kang16multiple}, two types of multiple-parameter DFRFT (MPDFRFT) were proposed, in which each spectral mode is assigned an independent fractional order. 

Let $\mathbf{D}_T$ be a $T \times T$ DFT matrix, with spectral decomposition $\mathbf{F}_T = \mathbf{V}_T\boldsymbol{\Lambda}_T\mathbf{V}_T^\mathrm{H}$, where $\mathbf{V}_T$ is the matrix of eigenvectors and $\boldsymbol{\Lambda}_T=\mathrm{diag}(\mu_0,\dots,\mu_{T-1})$ is the corresponding eigenvalue matrix. The type-I MPDFRFT (MPDFRFT-I) matrix with an order vector $\boldsymbol{a} \in \mathbb{R}^{T}$ is expressed as
\begin{equation}
	\mathbf{D}^{\boldsymbol{a}}_{\mathrm{I}}=\mathbf{V}_{T}\mathrm{diag}\left(\mu_0^{a_0},\dots,\mu_{T-1}^{a_{T-1}}\right)\mathbf{V}_T^{-1}.
\end{equation}

\indent The type-II MPDFRFT (MPDFRFT-II) matrix is defined as a linear combination of powers of $\mathbf{K}={\mathbf{D}_T}^{\frac{4}{T}}$
\begin{equation}
	\mathbf{D}^{\boldsymbol{a}}_{\mathrm{II}}=\sum_{t=0}^{T-1}D_{t,a_t}^{\mathrm{II}}\mathbf{K}^{t},
\end{equation}
where the coefficients $D_{t,a_t}^{\mathrm{II}}$ are 
\begin{align}
	D_{t,a_t}^{\mathrm{II}}=\frac{1}{T}\frac{1-\mathrm{e}^{\mathrm{j}2\pi\left(t-\frac{T}{4}a_t\right)}}{1-\mathrm{e}^{\mathrm{j}\frac{2\pi}{T}\left(t-\frac{T}{4}a_t\right)}}.
\end{align}

\section{Multiple-Parameter Joint Time-Vertex Fractional Fourier Transform} \label{sec3}
\subsection{Definitions}
\indent The JFRFT employs a single fractional order $\alpha$ for the GFRFT and a single order $\beta$ for the DFRFT. While this formulation provides a joint spectral representation of time-varying graph signals, it is restricted to a one-parameter family along each dimension, thereby limiting the flexibility of the transform. Consequently, the JFRFT may be insufficient to capture the complex spectral characteristics that evolve over time in graph-structured data.

\indent To overcome this limitation, we can extend the JFRFT by introducing multiple-parameter fractional orders in both the vertex and temporal dimensions. This enables the transform to adaptively characterize distinct spectral behaviors at different time instants by applying different MPGFRFTs to the graph signals corresponding to each moment. Depending on the choice of type-I or type-II operators for the vertex and temporal dimensions, this construction naturally leads to four variants of the DMPJFRFT.

\indent Consider a time-varying graph signal $\mathbf{X} = [\mathbf{x}_1, \dots, \mathbf{x}_T] \in \mathbb{C}^{N \times T}$. Let $\mathbf{A} = [\boldsymbol{a}^{(1)}, \boldsymbol{a}^{(2)}, \dots, \boldsymbol{a}^{(T)}] \in \mathbb{R}^{N \times T}$ denote the matrix of MPGFRFT orders, where $\boldsymbol{a}^{(i)} = (a^{(i)}_0, a^{(i)}_1, \dots, a^{(i)}_{N-1})^{\mathrm{T}} \in \mathbb{R}^{N}$ is the order vector applied to the $i$-th column $\mathbf{x}_i$ of $\mathbf{X}$ for $i=1,2,\dots,T$. Similarly, let $\boldsymbol{b} = (b_0, b_1, \dots, b_{T-1})^{\mathrm{T}} \in \mathbb{R}^{T}$ be the fractional order vector associated with the MPDFRFT along the temporal dimension. For these fractional order parameters $(\mathbf{A}, \boldsymbol{b})$, we can define four types of DMPJFRFT.

\indent \emph{Definition 1:} The type-I-I DMPJFRFT (DMPJFRFT-I-I) of $\mathbf{X}$ is defined as
\begin{align}
	&\mathrm{DMPJFRFT}_{\mathrm{I},\mathrm{I}}^{(\mathbf{A}, \boldsymbol{b})}(\mathbf{X}) \nonumber\\
	=&\left[\mathbf{F}_{\mathrm{I}}^{\boldsymbol{a}^{(1)}}\mathbf{x}_1, \mathbf{F}_{\mathrm{I}}^{\boldsymbol{a}^{(2)}}\mathbf{x}_2, \cdots, \mathbf{F}_{\mathrm{I}}^{\boldsymbol{a}^{(T)}}\mathbf{x}_T\right]\left(\mathbf{D}_{\mathrm{I}}^{\boldsymbol{b}}\right)^{\mathrm{T}}.
\end{align}

\indent \emph{Definition 2:} The type-I-II DMPJFRFT (DMPJFRFT-I-II) of $\mathbf{X}$ is defined as
\begin{align}
	=&\left[\mathbf{F}_{\mathrm{I}}^{\boldsymbol{a}^{(1)}}\mathbf{x}_1, \mathbf{F}_{\mathrm{I}}^{\boldsymbol{a}^{(2)}}\mathbf{x}_2, \cdots, \mathbf{F}_{\mathrm{I}}^{\boldsymbol{a}^{(T)}}\mathbf{x}_T\right]\left(\mathbf{D}_{\mathrm{II}}^{\boldsymbol{b}}\right)^{\mathrm{T}}.
\end{align}

\indent \emph{Definition 3:} The type-II-I DMPJFRFT (DMPJFRFT-II-I) of $\mathbf{X}$ is defined as
\begin{align}
	&\mathrm{DMPJFRFT}_{\mathrm{II},\mathrm{I}}^{(\mathbf{A}, \boldsymbol{b})}(\mathbf{X})\nonumber\\
	=&\left[\mathbf{F}_{\mathrm{II}}^{\boldsymbol{a}^{(1)}}\mathbf{x}_1, \mathbf{F}_{\mathrm{II}}^{\boldsymbol{a}^{(2)}}\mathbf{x}_2, \cdots, \mathbf{F}_{\mathrm{II}}^{\boldsymbol{a}^{(T)}}\mathbf{x}_T\right]\left(\mathbf{D}_{\mathrm{I}}^{\boldsymbol{b}}\right)^{\mathrm{T}}.
\end{align}

\indent \emph{Definition 4:} The type-II-II DMPJFRFT (DMPJFRFT-II-II) of $\mathbf{X}$ is defined as
\begin{align}
	&\mathrm{DMPJFRFT}_{\mathrm{II},\mathrm{II}}^{(\mathbf{A}, \boldsymbol{b})}(\mathbf{X})\nonumber\\
	=&\left[\mathbf{F}_{\mathrm{II}}^{\boldsymbol{a}^{(1)}}\mathbf{x}_1, \mathbf{F}_{\mathrm{II}}^{\boldsymbol{a}^{(2)}}\mathbf{x}_2, \cdots, \mathbf{F}_{\mathrm{II}}^{\boldsymbol{a}^{(T)}}\mathbf{x}_T\right]\left(\mathbf{D}_{\mathrm{II}}^{\boldsymbol{b}}\right)^{\mathrm{T}}.
\end{align}

\indent Let \(\mathbf{x}=\mathrm{vec}(\mathbf{X})\in\mathbb{C}^{NT}\) be the column-stacked vectorization of \(\mathbf{X}\). Define the block-diagonal MPGFRFT matrix as
\begin{align}
	\mathbf{F}_{\mathrm{blk},G_{\mathrm{type}}}^{\mathbf{A}}
	=\mathrm{blkdiag}\left(
	\mathbf{F}_{G_{\mathrm{type}}}^{\boldsymbol{a}^{(1)}},
	\mathbf{F}_{G_{\mathrm{type}}}^{\boldsymbol{a}^{(2)}},
	\cdots,
	\mathbf{F}_{G_{\mathrm{type}}}^{\boldsymbol{a}^{(T)}}
	\right) 
\end{align}
where $G_{\mathrm{type}}$ specifies the MPGFRFT type (I or II), and $\mathbf{F}_{\mathrm{blk},G_{\mathrm{type}}}^{\mathbf{A}}\in\mathbb{C}^{NT\times NT}$. Then, the four types of DMPJFRFT can be equivalently written in compact vectorized form as
\begin{align}
	\mathrm{DMPJFRFT}_{G_{\mathrm{type}},D_{\mathrm{type}}}^{(\mathbf{A}, \boldsymbol{b})}(\mathbf{x})
	= \mathbf{F}_{J,G_{\mathrm{type}},D_{\mathrm{type}}}^{(\mathbf{A}, \boldsymbol{b})}\,\mathbf{x},
\end{align}
where 
\begin{align} \label{DMPJFRFT matrix}
	\mathbf{F}_{J,G_{\mathrm{type}},D_{\mathrm{type}}}^{(\mathbf{A}, \boldsymbol{b})}
	=\left(\mathbf{D}_{D_{\mathrm{type}}}^{\boldsymbol{b}} \otimes \mathbf{I}_N\right)\mathbf{F}_{\mathrm{blk},G_{\mathrm{type}}}^{\mathbf{A}}
\end{align}
is the DMPJFRFT matrix, and $D_{\mathrm{type}}$ denotes the chosen MPDFRFT type (I or II).

\begin{figure*}
	\centering
	\includegraphics[width=6in]{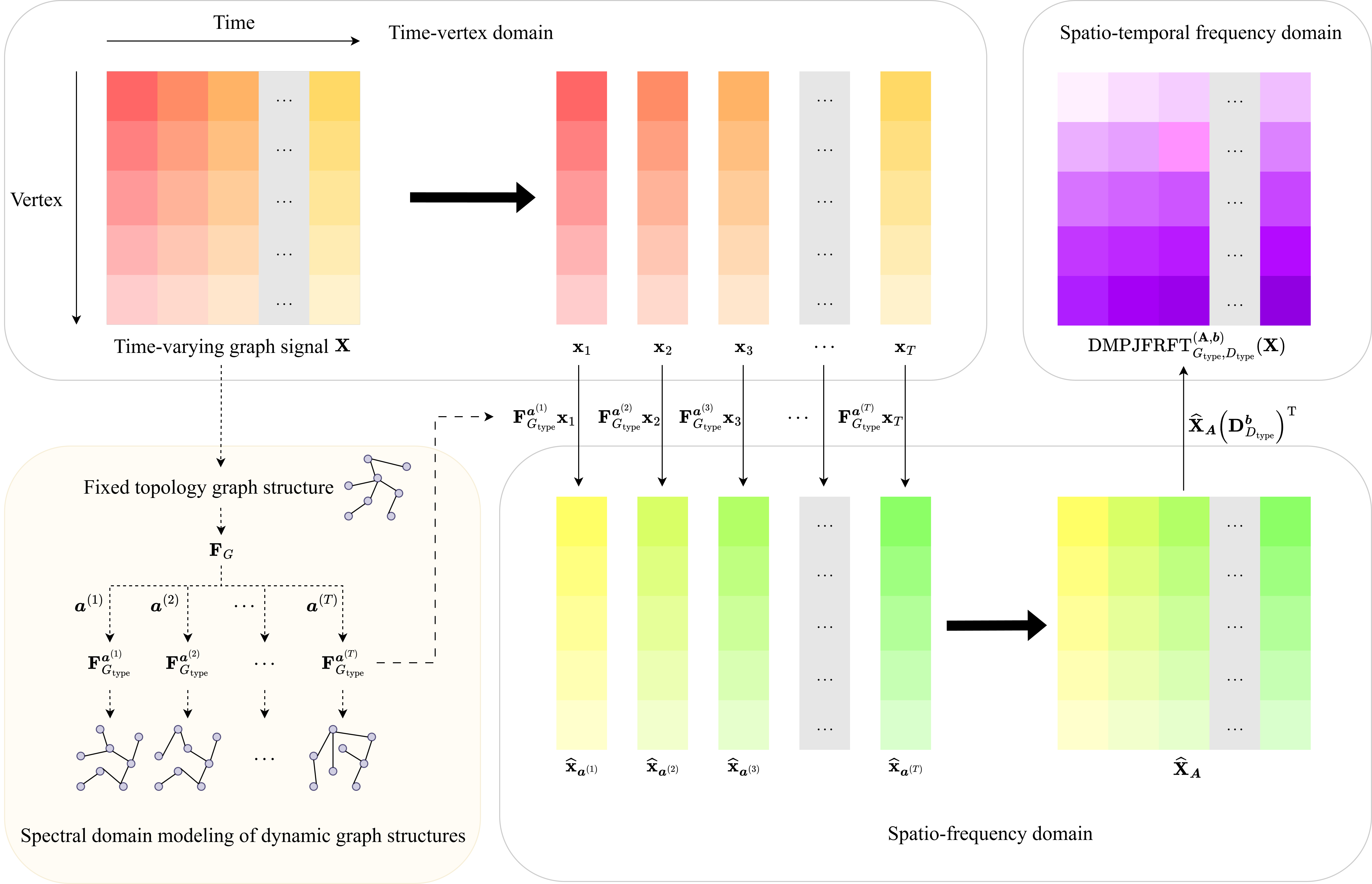}
	\caption{Conceptual illustration of the DMPJFRFT mechanism.}
	\label{fig:DMPJFRFT_concept}
\end{figure*}

\indent \emph{Remark 1:} It is worth noting that the proposed DMPJFRFT framework is not merely a multiple-parameter extension of the conventional JFRFT. Its core idea is to apply a distinct set of fractional parameters to each time step of a dynamic graph signal, taking into account the temporal variations inherent in dynamic graph signals, and effectively modeling the signal in the graph spectral domain at each moment. By using different fractional parameters across time, the underlying graph spectral representation evolves dynamically, capturing temporal changes in the graph structure. In this way, DMPJFRFT achieves adaptive time-varying graph structure modeling in the multiple-parameter spectral domain, providing a faithful and flexible representation of dynamic graph signals. To provide an intuitive understanding, Fig.~\ref{fig:DMPJFRFT_concept} illustrates the overall mechanism of the proposed DMPJFRFT framework.

\subsection{Properties}
\indent \emph{Property 1 (Identity transform):} If the fractional order parameters are set to $(\mathbf{A}, \boldsymbol{b})=(\mathbf{0}_{N\times T},\mathbf{0}_{T\times 1})$, then all four types of DMPJFRFT reduce to the identity transform.

\indent \emph{Property 2 (Reduction to JFRFT):} If the fractional order parameters are set to $(\mathbf{A}, \boldsymbol{b})=(\boldsymbol{a}_{N\times T},\mathbf{b}_{T\times 1})$, then all four types of DMPJFRFT reduce to the JFRFT.

\indent \emph{Property 3 (Reduction to JFT):} If the fractional order parameters are set to $(\mathbf{A}, \boldsymbol{b})=(\mathbf{1}_{N\times T},\mathbf{1}_{T\times 1})$, then all four types of DMPJFRFT reduce to the JFT.

\indent \emph{Proof:} The proofs of properties 1--3 are straightforward and thus omitted here.

\indent \emph{Property 4 (Index additivity):} Let $(\mathbf{A}_1, \boldsymbol{b}_1)$ and $(\mathbf{A}_2, \boldsymbol{b}_2)$ be two sets of fractional order parameters, where $\mathbf{A}_1 = [\boldsymbol{a}_1^{(1)}, \boldsymbol{a}_1^{(2)}, \dots, \boldsymbol{a}_1^{(T)}]$ and $\mathbf{A}_2 = [\boldsymbol{a}_2^{(1)}, \boldsymbol{a}_2^{(2)}, \dots, \boldsymbol{a}_2^{(T)}]$. If the MPGFRFT orders in $\mathbf{A}_2$ are time-invariant, i.e. $\boldsymbol{a}_2^{(1)}=\boldsymbol{a}_2^{(2)}=\cdots=\boldsymbol{a}_2^{(T)}$, and both the MPGFRFT matrices $\mathbf{F}_{G_{\mathrm{type}}}^{\boldsymbol{a}_r^{(i)}}$ and the MPDFRFT matrices $\mathbf{D}_{D_{\mathrm{type}}}^{\boldsymbol{b}_r}$ $(r=1,2)$ satisfy the index additivity property, then the DMPJFRFT satisfies the additivity property that
\begin{align}
	&\mathrm{DMPJFRFT}_{G_{\mathrm{type}},D_{\mathrm{type}}}^{(\mathbf{A}_2, \boldsymbol{b}_2)}\left(\mathrm{DMPJFRFT}_{G_{\mathrm{type}},D_{\mathrm{type}}}^{(\mathbf{A}_1, \boldsymbol{b}_1)}(\mathbf{X})\right) \nonumber\\
	=&\mathrm{DMPJFRFT}_{G_{\mathrm{type}},D_{\mathrm{type}}}^{(\mathbf{A}_1 + \mathbf{A}_2, \boldsymbol{b}_1 + \boldsymbol{b}_2)}(\mathbf{X}).
\end{align}
\indent \emph{Proof:} According to \eqref{DMPJFRFT matrix}, we have
\begin{align}
	&\mathbf{F}_{J,G_{\mathrm{type}},D_{\mathrm{type}}}^{(\mathbf{A}_2, \boldsymbol{b}_2)}\mathbf{F}_{J,G_{\mathrm{type}},D_{\mathrm{type}}}^{(\mathbf{A}_1, \boldsymbol{b}_1)} \nonumber\\
	=&\left(\mathbf{D}_{D_{\mathrm{type}}}^{\boldsymbol{b}_2} \otimes \mathbf{I}_N\right)\mathbf{F}_{\mathrm{blk},G_{\mathrm{type}}}^{\mathbf{A}_2}
	\left(\mathbf{D}_{D_{\mathrm{type}}}^{\boldsymbol{b}_1} \otimes \mathbf{I}_N\right)\mathbf{F}_{\mathrm{blk},G_{\mathrm{type}}}^{\mathbf{A}_1}.
\end{align}

If $\boldsymbol{a}_2^{(1)}=\boldsymbol{a}_2^{(2)}=\cdots=\boldsymbol{a}_2^{(T)}$, the above equation can be written as
\begin{align}
	&\mathbf{F}_{J,G_{\mathrm{type}},D_{\mathrm{type}}}^{(\mathbf{A}_2, \boldsymbol{b}_2)}\mathbf{F}_{J,G_{\mathrm{type}},D_{\mathrm{type}}}^{(\mathbf{A}_1, \boldsymbol{b}_1)} \nonumber\\
	=&\left(\mathbf{D}_{D_{\mathrm{type}}}^{\boldsymbol{b}_2} \otimes \mathbf{I}_N\right)
	\left(\mathbf{D}_{D_{\mathrm{type}}}^{\boldsymbol{b}_1} \otimes \mathbf{I}_N\right)
	\mathbf{F}_{\mathrm{blk},G_{\mathrm{type}}}^{\mathbf{A}_2}\mathbf{F}_{\mathrm{blk},G_{\mathrm{type}}}^{\mathbf{A}_1} \nonumber\\
	=&\left(\mathbf{D}_{D_{\mathrm{type}}}^{\boldsymbol{b}_2}\mathbf{D}_{D_{\mathrm{type}}}^{\boldsymbol{b}_1} \otimes \mathbf{I}_N\right)
	\mathbf{F}_{\mathrm{blk},G_{\mathrm{type}}}^{\mathbf{A}_2}\mathbf{F}_{\mathrm{blk},G_{\mathrm{type}}}^{\mathbf{A}_1}.
\end{align}
\indent It is obvious that
\begin{align}
	&\mathbf{F}_{\mathrm{blk},G_{\mathrm{type}}}^{\mathbf{A}_2}\mathbf{F}_{\mathrm{blk},G_{\mathrm{type}}}^{\mathbf{A}_1} \nonumber\\
	=&\mathrm{blkdiag}\left(
	\mathbf{F}_{G_{\mathrm{type}}}^{\boldsymbol{a}_2^{(1)}}\mathbf{F}_{G_{\mathrm{type}}}^{\boldsymbol{a}_1^{(1)}},
	\mathbf{F}_{G_{\mathrm{type}}}^{\boldsymbol{a}_2^{(2)}}\mathbf{F}_{G_{\mathrm{type}}}^{\boldsymbol{a}_1^{(2)}},
	\cdots,
	\mathbf{F}_{G_{\mathrm{type}}}^{\boldsymbol{a}_2^{(T)}}\mathbf{F}_{G_{\mathrm{type}}}^{\boldsymbol{a}_1^{(T)}}
	\right),
\end{align}
thus, if both the MPGFRFT matrices $\mathbf{F}_{G_{\mathrm{type}}}^{\boldsymbol{a}_r^{(i)}}$ and the MPDFRFT matrices $\mathbf{D}_{D_{\mathrm{type}}}^{\boldsymbol{b}_r}$ $(r=1,2)$ satisfy the index additivity property, we arrive the required result. \qed

\indent \emph{Property 5 (Reversibility):} If each MPGFRFT matrix $\mathbf{F}_{G_{\mathrm{type}}}^{\boldsymbol{a}^{(i)}}$ $(i=1,2,\cdots,T)$ and the MPDFRFT matrix $\mathbf{D}_{D_{\mathrm{type}}}^{\boldsymbol{b}}$ are invertible, the DMPJFRFT is invertible. The inverse DMPJFRFT matrix is given by
\begin{align}
	\left(\mathbf{F}_{J,G_{\mathrm{type}},D_{\mathrm{type}}}^{(\mathbf{A},\boldsymbol{b})}\right)^{-1}
	=\left(\mathbf{F}_{\mathrm{blk},G_{\mathrm{type}}}^{\mathbf{A}}\right)^{-1} \left( \left(\mathbf{D}_{D_{\mathrm{type}}}^{\boldsymbol{b}}\right)^{-1} \otimes \mathbf{I}_N\right).
\end{align}
In particular, for the DMPJFRFT-I-I case, we have
\begin{align}
	\left(\mathbf{F}_{J,\mathrm{I},\mathrm{I}}^{(\mathbf{A},\boldsymbol{b})}\right)^{-1}
	=\mathbf{F}_{J,\mathrm{I},\mathrm{I}}^{(-\mathbf{A},-\boldsymbol{b})}.
\end{align}
\indent \emph{Proof:} If each MPGFRFT matrix $\mathbf{F}_{G_{\mathrm{type}}}^{\boldsymbol{a}^{(i)}}$ $(i=1,2,\cdots,T)$ are invertible, the block-diagonal MPGFRFT matrix is invertible
\begin{align}
	\left(\mathbf{F}_{\mathrm{blk},G_{\mathrm{type}}}^{\mathbf{A}}\right)^{-1}
	=\mathrm{blkdiag}\left(
	\left( \mathbf{F}_{G_{\mathrm{type}}}^{\boldsymbol{a}^{(1)}}\right)^{-1} ,
	\cdots,
	\left( \mathbf{F}_{G_{\mathrm{type}}}^{\boldsymbol{a}^{(T)}}\right)^{-1}
	\right).
\end{align}
\indent If the MPDFRFT matrix $\mathbf{D}_{D_{\mathrm{type}}}^{\boldsymbol{b}}$ is invertible, we have
\begin{align}
	\left(\mathbf{D}_{D_{\mathrm{type}}}^{\boldsymbol{b}} \otimes \mathbf{I}_N\right)^{-1}
	=\left(\mathbf{D}_{D_{\mathrm{type}}}^{\boldsymbol{b}}\right)^{-1} \otimes \mathbf{I}_N.
\end{align}
\indent In particular, if $G_{\mathrm{type}}=\mathrm{I}$ and $D_{\mathrm{type}}=\mathrm{I}$, we have
\begin{align}
	\left(\mathbf{F}_{\mathrm{blk},G_{\mathrm{type}}}^{\mathbf{A}}\right)^{-1}
	=&\mathrm{blkdiag}\left(
	\mathbf{F}_{G_{\mathrm{type}}}^{-\boldsymbol{a}^{(1)}} ,
	\cdots,\mathbf{F}_{G_{\mathrm{type}}}^{-\boldsymbol{a}^{(T)}}
	\right) \nonumber\\
	=&\mathbf{F}_{\mathrm{blk},G_{\mathrm{type}}}^{-\mathbf{A}},
\end{align}
\begin{align}
	\left(\mathbf{D}_{D_{\mathrm{type}}}^{\boldsymbol{b}} \otimes \mathbf{I}_N\right)^{-1}
	=\mathbf{D}_{D_{\mathrm{type}}}^{-\boldsymbol{b}} \otimes \mathbf{I}_N.
\end{align}
\indent Thus,  we arrive the required results.  \qed

\indent \emph{Property 6 (Index commutativity):} Let $(\mathbf{A}_1, \boldsymbol{b}_1)$ and $(\mathbf{A}_2, \boldsymbol{b}_2)$ be two sets of fractional order parameters.
If the MPDFRFT matrices are commutative, i.e. $\mathbf{D}_{D_{\mathrm{type}}}^{\boldsymbol{b}_1}\mathbf{D}_{D_{\mathrm{type}}}^{\boldsymbol{b}_2}=\mathbf{D}_{D_{\mathrm{type}}}^{\boldsymbol{b}_2}\mathbf{D}_{D_{\mathrm{type}}}^{\boldsymbol{b}_1}$, the MPGFRFT orders in $\mathbf{A}_1$ and $\mathbf{A}_2$ are time-invariant and the single-block MPGFRFT matrices are commutative, i.e.
$\mathbf{F}_{G_{\mathrm{type}}}^{\boldsymbol{a}_1^{(1)}}\mathbf{F}_{G_{\mathrm{type}}}^{\boldsymbol{a}_2^{(1)}}=\mathbf{F}_{G_{\mathrm{type}}}^{\boldsymbol{a}_2^{(1)}}\mathbf{F}_{G_{\mathrm{type}}}^{\boldsymbol{a}_1^{(1)}}$, 
then the corresponding DMPJFRFT satisfies the commutativity property that
\begin{align}
	&\mathrm{DMPJFRFT}_{G_{\mathrm{type}},D_{\mathrm{type}}}^{(\mathbf{A}_2, \boldsymbol{b}_2)}
	\mathrm{DMPJFRFT}_{G_{\mathrm{type}},D_{\mathrm{type}}}^{(\mathbf{A}_1, \boldsymbol{b}_1)} \nonumber\\
	=&\mathrm{DMPJFRFT}_{G_{\mathrm{type}},D_{\mathrm{type}}}^{(\mathbf{A}_1, \boldsymbol{b}_1)}
	\mathrm{DMPJFRFT}_{G_{\mathrm{type}},D_{\mathrm{type}}}^{(\mathbf{A}_2, \boldsymbol{b}_2)}.
\end{align}

\indent \emph{Proof:} Under the time-invariance assumption,  each block-diagonal matrix can be written in Kronecker form as
\begin{align}
	\mathbf{F}_{\mathrm{blk},G_{\mathrm{type}}}^{\mathbf{A}_1}
	= \mathbf{I}_T\otimes\mathbf{F}_{G_{\mathrm{type}}}^{\boldsymbol{a}_1^{(1)}},\quad
	\mathbf{F}_{\mathrm{blk},G_{\mathrm{type}}}^{\mathbf{A}_2}
	= \mathbf{I}_T\otimes\mathbf{F}_{G_{\mathrm{type}}}^{\boldsymbol{a}_2^{(1)}}.
\end{align}
\indent Then, we have
\begin{align}
	&\mathbf{F}_{J}^{(\mathbf{A}_2,\boldsymbol{b}_2)}\,
	\mathbf{F}_{J}^{(\mathbf{A}_1,\boldsymbol{b}_1)} \nonumber\\
	=&\left( \mathbf{D}_{D_{\mathrm{type}}}^{\boldsymbol{b}_2}\otimes\mathbf{I}_N\right) \left( \mathbf{I}_T\otimes\mathbf{F}_{G}^{\boldsymbol{a}_2^{(1)}}\right) 
	\left( \mathbf{D}_{D_{\mathrm{type}}}^{\boldsymbol{b}_1}\otimes\mathbf{I}_N\right) \nonumber\\
	&\times\left( \mathbf{I}_T\otimes\mathbf{F}_{G}^{\boldsymbol{a}_1^{(1)}}\right)  \nonumber\\
	=&\left(\mathbf{D}_{D_{\mathrm{type}}}^{\boldsymbol{b}_2}\mathbf{D}_{D_{\mathrm{type}}}^{\boldsymbol{b}_1}\right) \otimes
	\left( \mathbf{F}_{G}^{\boldsymbol{a}_2^{(1)}}\mathbf{F}_{G}^{\boldsymbol{a}_1^{(1)}}\right) .
\end{align}
\indent Since the MPDFRFT and single-block MPGFRFT matrices are assumed to be commutative, i.e.,
$\mathbf{D}_{D_{\mathrm{type}}}^{\boldsymbol{b}_1}\mathbf{D}_{D_{\mathrm{type}}}^{\boldsymbol{b}_2}=\mathbf{D}_{D_{\mathrm{type}}}^{\boldsymbol{b}_2}\mathbf{D}_{D_{\mathrm{type}}}^{\boldsymbol{b}_1}$,
\begin{align}
	\mathbf{D}_{D_{\mathrm{type}}}^{\boldsymbol{b}_1}\mathbf{D}_{D_{\mathrm{type}}}^{\boldsymbol{b}_2}=\mathbf{D}_{D_{\mathrm{type}}}^{\boldsymbol{b}_2}\mathbf{D}_{D_{\mathrm{type}}}^{\boldsymbol{b}_1},
\end{align}
\begin{align}
	\mathbf{F}_{G}^{\boldsymbol{a}_2^{(1)}}\mathbf{F}_{G}^{\boldsymbol{a}_1^{(1)}}=\mathbf{F}_{G}^{\boldsymbol{a}_1^{(1)}}\mathbf{F}_{G}^{\boldsymbol{a}_2^{(1)}},
\end{align}
we arrive the required result. \qed

\indent \emph{Property 7 (Linearity):} Let $\mathbf{X}, \mathbf{Y} \in \mathbb{C}^{N \times T}$ be two time-varying graph signals and $\alpha, \beta \in \mathbb{C}$. The DMPJFRFT is linear, i.e.,
\begin{align}
	&\mathrm{DMPJFRFT}_{G_{\mathrm{type}},D_{\mathrm{type}}}^{(\mathbf{A}, \boldsymbol{b})}
	\left(\alpha \mathbf{X} + \beta \mathbf{Y} \right) \nonumber\\
	=& \alpha\mathrm{DMPJFRFT}_{G_{\mathrm{type}},D_{\mathrm{type}}}^{(\mathbf{A}, \boldsymbol{b})}\left(\mathbf{X}\right) \nonumber\\
	&+ \beta\mathrm{DMPJFRFT}_{G_{\mathrm{type}},D_{\mathrm{type}}}^{(\mathbf{A}, \boldsymbol{b})}
	\left(\mathbf{Y}\right).
\end{align}
\indent \emph{Proof:} Let $\mathbf{x} = \mathrm{vec}(\mathbf{X})$ and $\mathbf{y} = \mathrm{vec}(\mathbf{Y})$. Then, we have
\begin{align}
	&\mathrm{DMPJFRFT}_{G_{\mathrm{type}},D_{\mathrm{type}}}^{(\mathbf{A}, \boldsymbol{b})}
	\left(\alpha \mathbf{X} + \beta \mathbf{Y} \right)\nonumber\\
	=&\mathbf{F}_{J,G_{\mathrm{type}},D_{\mathrm{type}}}^{(\mathbf{A}, \boldsymbol{b})}
	\left(\alpha \mathbf{x} + \beta \mathbf{y}\right) \nonumber\\
	=& \alpha \mathbf{F}_{J,G_{\mathrm{type}},D_{\mathrm{type}}}^{(\mathbf{A}, \boldsymbol{b})}\mathbf{x}
	+ \beta \mathbf{F}_{J,G_{\mathrm{type}},D_{\mathrm{type}}}^{(\mathbf{A}, \boldsymbol{b})}\mathbf{y}.
\end{align}
Thus, we arrive the required result. \qed

\subsection{Computational complexity}

For the the GFRFT and MPGFRFT-I, the main computational cost arises from the spectral decomposition or the computation of fractional powers of the GFT matrix, resulting in a complexity of $\mathcal{O}\left(N^{3}\right)$. For the MPGFRFT-II matrix, the matrix powers $\mathbf{F}_G^{n}$ can be precomputed, thereby reducing repeated computation during multiple transformations. Thus, the computational complexity of constructing the MPGFRFT-II matrix remains $\mathcal{O}\left(N^{3}\right)$. Similarly, the DFRFT and MPDFRFT in the temporal domain require the spectral decomposition or the computation of fractional powers of the DFT matrix, leading to a computational complexity of $\mathcal{O}\left(T^{3}\right)$. 

\indent For an $N \times T$ time-varying graph signal $\mathbf{X}$, the computational cost of the JFRFT mainly arises from two matrix multiplications involving the graph-domain and temporal-domain fractional transform matrices. Assuming these matrices are precomputed, the overall per-transformation complexity is $\mathcal{O}\left(N^{2}T + NT^{2}\right)$. When $N \approx T$, this complexity can be approximated as $\mathcal{O}\left(N^{3}\right)$. Therefore, the DMPJFRFT has the same complexity as the conventional JFRFT while providing enhanced spectral flexibility through multi-parameter fractional orders. Table~\ref{tab:complexity} gives a summary of the computational complexity of the proposed transforms.

\begin{table}[htbp]
	\centering
	\caption{Computational complexity of different fractional transforms}
	\begin{tabular}{lcc}
		\toprule
		\textbf{Transform} & \textbf{Domain} & \textbf{Complexity} \\
		\midrule
		GFRFT & Graph spectral & $\mathcal{O}(N^{3})$ \\
		MPGFRFT-I & Graph spectral  & $\mathcal{O}(N^{3})$ \\
		MPGFRFT-II & Graph spectral & $\mathcal{O}(N^{3})$ \\
		\midrule
		DFRFT & Temporal & $\mathcal{O}(T^{3})$ \\
		MPDFRFT-I & Temporal & $\mathcal{O}(T^{3})$ \\
		MPDFRFT-II & Temporal & $\mathcal{O}(T^{3})$ \\
		\midrule
		JFRFT & Joint & $\mathcal{O}(N^{2}T + NT^{2})$ \\
		DMPJFRFT & Joint & $\mathcal{O}(N^{2}T + NT^{2})$ \\
		\bottomrule
	\end{tabular}
	\label{tab:complexity}
\end{table}

\section{Gradient Descent-Based DMPJFRFT Filtering}  \label{sec4}
\indent The multiple-parameter mechanism enables fine-grained manipulation of the signal representation, which is particularly beneficial for problems such as denoising and deblurring. In this Section, we develop gradient descent-based filtering schemes in the DMPJFRFT domain.

\indent In this work, we consider two typical signal degradation models: denoising and deblurring, both of which aim to recover the clean time-varying graph signal from its corrupted observation. For the denoising task, the degradation model is given by $\mathbf{Y}=\mathbf{X}+\mathbf{N}$, where $\mathbf{X}\in \mathbb{C}^{N\times T}$ is the pure time-varying graph signal, $\mathbf{Y}$ is the observed corrupted signal, and $\mathbf{N}$ represents additive noise. For the deblurring task, the degradation process can be modeled as $\mathbf{Y}=\mathbf{KX}$, where $\mathbf{K}$ denotes the blurring operator. Regardless of whether the corruption arises from additive noise or blurring, the filtering process in both denoising and deblurring scenarios aims to recover $\mathbf{X}$ from $\mathbf{Y}$ by minimizing the mean square error (MSE)
\begin{align}
	\mathcal{L}_{\mathrm{MSE}} = \frac{1}{NT}\|\widetilde{\mathbf{X}}- \mathbf{X}\|_F^2,
\end{align}
where $\widetilde{\mathbf{X}}$ is the reconstructed signal after filtering. Specifically, the reconstruction process can be expressed as
\begin{align}
\mathrm{vec}\left( \widetilde{\mathbf{X}}\right)=\left( \mathbf{F}_{J,G_{\mathrm{type}},D_{\mathrm{type}}}^{(\mathbf{A}, \boldsymbol{b})}\right)^{-1}\mathbf{H} \mathbf{F}_{J,G_{\mathrm{type}},D_{\mathrm{type}}}^{(\mathbf{A}, \boldsymbol{b})}\,\mathbf{x},
\end{align} 
where $\mathbf{x}=\mathrm{vec}(\mathbf{X})$ and $\mathbf{H}$ is the learnable diagonal filter.

\indent Since both the MPGFRFT and MPDFRFT are differentiable with respect to their transform parameters, the DMPJFRFT naturally inherits this property and is therefore differentiable with respect to $(\mathbf{A}, \boldsymbol{b})$. Consequently, gradient descent can be effectively employed to jointly optimize the transform parameters and the spectral filter.

\indent The parameters are updated iteratively by
\begin{align}
	\left(\mathbf{A},\boldsymbol{b}\right)^{(t+1)}= \left(\mathbf{A},\boldsymbol{b}\right)^{(t)}-\gamma\left(  \frac{\partial \mathcal{L}}{\partial\mathbf{A}}, \frac{\partial \mathcal{L}}{\partial\boldsymbol{b}}\right),
\end{align}
\begin{align}
	\mathbf{H}^{(t+1)}= \mathbf{H}^{(t)}-\gamma \frac{\partial \mathcal{L}}{\partial\mathbf{H}},
\end{align}
where $\gamma$ is the learning rate.

\indent Note that for the MPGFRFT-II, the transformation involves the matrix defined $\mathbf{P}$ in \eqref{eq:MPGFRFT_P}. When the modulus of some eigenvalues deviates from one and the number of vertices $N$ is large, computing $\mathbf{P}$ may become numerically unstable or even fail due to ill-conditioning. Consequently, in the following experiments, we restrict our attention to DMPJFRFT-I-I and DMPJFRFT-I-II.

\indent To quantitatively evaluate the filtering performance, we adopt the signal-to-noise ratio (SNR), defined as
\begin{align}
	\mathrm{SNR} = 20 \log_{10} \frac{\|\mathbf{X}\|_F}{\|\widetilde{\mathbf{X}} - \mathbf{X}\|_F},
\end{align}
Higher SNR values indicate better recovery quality.

\subsection{Denoising}  \label{Gradient Descent Denoising}
\indent \emph{1) Real-data denoising:} For graph signal denoising, we use four real-world datasets: PEMSD7(M), PEMS08, Quality \cite{Guo22Learning,yan2025jfrffnet,Kun23Frequency}, and SST dataset \cite{Giraldo22}. For each dataset, the first 30 temporal slices are extracted, and every 10 consecutive columns are regarded as a single time-varying graph signal for filtering. In the experiments, the learning rate is set to $0.01$, and the number of training epochs is $1000$. The diagonal filter $\mathbf{H}$ is initialized as an identity matrix, while the transform parameters of the DMPJFRFT are all initialized to $0.5$.

\indent We compare the proposed DMPJFRFT-based filtering schemes with three baseline methods: two-dimensional GFRFT (2D GFRFT) \cite{YAN2022Multi}, 2D graph bi-fractional Fourier transform (2D GBFRFT), and JFRFT. All methods are evaluated under five types of GSOs: the adjacency matrix (adj), the Laplacian (lap), the normalized Laplacian (nor lap), the row-normalized adjacency matrix (row nor adj), and the symmetrically normalized adjacency matrix (sym nor adj). To further assess robustness, experiments are conducted under multiple noise levels controlled by the standard deviation parameter $\sigma$. Table~\ref{tab:Gradient Descent denoise signal} presents the SNR results of these methods across different GSOs and noise levels. The results clearly show that both types of DMPJFRFT-based filters exhibit consistently excellent denoising performance, surpassing all comparison methods across all experimental settings.

\begin{table*}[htbp]
	\caption{SNR performance of gradient descent-based denoising methods under various GSOs for real-world graph signals.
		\label{tab:Gradient Descent denoise signal}}
	\centering
	\resizebox{1.0\linewidth}{!}{
		\begin{tabular}{cccccccccccccccccc}
			\toprule[1.5pt]
		\multicolumn{1}{c}{\multirow{2}{*}{PEMSD7(M)}} & \multicolumn{5}{c}{$\sigma=70$} && \multicolumn{5}{c}{$\sigma=80$} && \multicolumn{5}{c}{$\sigma=90$} \\
		~ & \multicolumn{5}{c}{(SNR=$-0.446$)} && \multicolumn{5}{c}{(SNR=$-1.606$)} && \multicolumn{5}{c}{(SNR=$-2.629$)} \\
		\cmidrule{2-6} \cmidrule{8-12} \cmidrule{14-18}
			~ & 2D GFRFT & 2D GBFRFT & JFRFT & DMPJFRFT-I-I & DMPJFRFT-I-II && 2D GFRFT & 2D GBFRFT & JFRFT & DMPJFRFT-I-I & DMPJFRFT-I-II && 2D GFRFT & 2D GBFRFT & JFRFT & DMPJFRFT-I-I & DMPJFRFT-I-II \\ \midrule
			adj  & $21.813$  & $22.414$  & $22.246$  & $\mathbf{33.438}$  & $23.502$   & ~ & $21.128$  & $21.669$  & $21.516$  & $\mathbf{33.227}$  & $24.355$   & ~ & $20.463$  & $20.955$  & $20.816$  & $\mathbf{32.870}$  & $24.451$   \\ 
			lap  & $22.018$  & $22.411$  & $22.419$  & $\mathbf{33.330} $ & $25.788$   & ~ & $21.287$  & $21.678$  & $21.625$  & $\mathbf{33.363}$  & $25.503$   & ~ & $20.586$  & $20.981$  & $20.899$  & $\mathbf{33.053}$  & $24.349$   \\ 
			nor lap  & $21.888$  & $22.373$  & $22.204$  & $\mathbf{33.206}$  & $25.387$   & ~ & $21.180$  & $21.671$  & $21.487$  & $\mathbf{33.435}$  & $25.225$   & ~ & $20.499$  & $20.960$  & $20.841$  & $\mathbf{32.815}$  & $25.408$   \\ 
			row nor adj  & $21.854$  & $22.390$  & $22.228$  & $\mathbf{33.259}$  & $24.310$   & ~ & $21.158$  & $21.649$  & $21.503$  & $\mathbf{33.333}$  & $24.728$   & ~ & $20.481$  & $20.946$  & $20.808$  & $\mathbf{33.227}$  & $24.614$   \\ 
			sym nor adj  & $21.854$  & $22.390$  & $22.228$  & $\mathbf{33.259}$  & $24.310$   & ~ & $21.158$  & $21.649$  & $21.503$  & $\mathbf{33.333}$  & $24.728$   & ~ & $20.481$  & $20.946$  & $20.808$  & $\mathbf{33.227}$  & $24.614$   \\ 
			\bottomrule[1.5pt]
			\toprule[1.5pt]
		\multicolumn{1}{c}{\multirow{2}{*}{PEMS08}} & \multicolumn{5}{c}{$\sigma=80$} && \multicolumn{5}{c}{$\sigma=100$} && \multicolumn{5}{c}{$\sigma=120$} \\
		~ & \multicolumn{5}{c}{(SNR=$1.865$)} && \multicolumn{5}{c}{(SNR=$-0.073$)} && \multicolumn{5}{c}{(SNR=$-1.657$)} \\
		\cmidrule{2-6} \cmidrule{8-12} \cmidrule{14-18}
			adj & $8.284$  & $8.416$  & $14.294$  & $\mathbf{23.045}$  & $18.518$   & ~ & $7.190$  & $7.343$  & $13.164$  & $\mathbf{21.917}$  & $17.930$   & ~ & $6.357$  & $6.520$  & $12.121$  & $\mathbf{22.287}$  & $17.656$   \\ 
			lap & $9.995$  & $10.250$  & $14.249$  & $\mathbf{22.449}$  & $19.150$   & ~ & $9.277$  & $9.554$  & $13.101$  & $\mathbf{22.588}$  & $17.239$   & ~ & $8.764$  & $9.382$  & $12.289$  & $\mathbf{21.343}$  & $18.345$   \\ 
			nor lap & $9.659$  & $9.902$  & $14.501$  & $\mathbf{22.746}$  & $18.682$   & ~ & $8.793$  & $9.042$  & $13.417$  & $\mathbf{22.101}$  & $18.613$   & ~ & $8.159$  & $8.407$  & $12.525$  & $\mathbf{21.874}$  & $17.947$   \\ 
			row nor adj & $9.888$  & $10.025$  & $14.563$  & $\mathbf{23.085}$  & $17.648$   & ~ & $9.161$  & $9.311$  & $13.498$  & $\mathbf{22.699}$  & $17.072$   & ~ & $8.628$  & $8.813$  & $12.613$  & $\mathbf{22.361}$  & $17.082 $  \\ 
			sym nor adj  & $9.717$  & $9.926$  & $14.523$  & $\mathbf{23.030}$  & $18.138$   & ~ & $8.845$  & $9.063$  & $13.430$  & $\mathbf{22.809}$  & $18.116$   & ~ & $8.195$  & $8.426$  & $12.524$  & $\mathbf{22.397}$  & $18.703$   \\ 
			\bottomrule[1.5pt]
			\toprule[1.5pt]
		\multicolumn{1}{c}{\multirow{2}{*}{Quality}} & \multicolumn{5}{c}{$\sigma=80$} && \multicolumn{5}{c}{$\sigma=90$} && \multicolumn{5}{c}{$\sigma=100$} \\
		~ & \multicolumn{5}{c}{(SNR=$2.034$)} && \multicolumn{5}{c}{(SNR=$1.011$)} && \multicolumn{5}{c}{(SNR=$0.095$)} \\
		\cmidrule{2-6} \cmidrule{8-12} \cmidrule{14-18}
			adj & $6.776$  & $7.024$  & $12.608$  & $\mathbf{20.540}$  & $17.693$   & ~ & $6.127$  & $6.374$  & $11.868$  & $\mathbf{20.984}$  & $16.648$ & ~ & $5.575$  & $5.803$  & $11.187$  & $\mathbf{20.645}$  & $15.897$   \\ 
			lap & $6.913$  & $7.239$  & $12.965$  & $\mathbf{20.604}$  & $15.882$   & ~ & $6.213$  & $6.578$  & $12.328$  & $\mathbf{21.109}$  & $15.069$ & ~ & $5.619$  & $6.010$  & $11.754$  & $\mathbf{21.369}$  & $15.316$ \\ 
			nor lap & $6.684$  & $6.925$  & $14.133$  & $\mathbf{21.429}$  & $15.636$   & ~ & $6.124$  & $6.268$  & $13.594$  & $\mathbf{21.664}$  & $15.163$ & ~ & $5.550$  & $6.123$  & $13.088$  & $\mathbf{21.617}$  & $16.056$  \\ 
			row nor adj & $7.552$  & $7.593$  & $12.708$  & $16.795$  & $\mathbf{16.911}$   & ~ & $6.862$  & $9.644$  & $12.001$  & $16.121$  & $\mathbf{17.737}$ & ~ & $6.263$  & $9.144$  & $11.353$  & $15.866$  & $\mathbf{19.120}$   \\ 
			sym nor adj & $6.234$  & $6.534$  & $12.585$  & $\mathbf{21.381}$  & $15.513$   & ~ & $5.613$  & $6.163$  & $11.943$  & $\mathbf{21.370}$  & $16.339$ & ~ & $5.559$  & $9.221$  & $11.292$  & $\mathbf{21.695}$  & $17.887$ \\ 
			\bottomrule[1.5pt]
			\toprule[1.5pt]
		\multicolumn{1}{c}{\multirow{2}{*}{SST}} & \multicolumn{5}{c}{$\sigma=15$} && \multicolumn{5}{c}{$\sigma=20$} && \multicolumn{5}{c}{$\sigma=25$} \\
		~ & \multicolumn{5}{c}{(SNR=$2.987$)} && \multicolumn{5}{c}{(SNR=$0.488$)} && \multicolumn{5}{c}{(SNR=$-1.450$)} \\
		\cmidrule{2-6} \cmidrule{8-12} \cmidrule{14-18}
			adj & $12.624$  & $12.793$  & $18.287$  & $\mathbf{29.720}$  & $19.467$   & ~ & $11.172$  & $11.377$  & $16.804$  & $\mathbf{29.075}$  & $19.595$   & ~ & $10.105$  & $10.271$  & $15.600$  & $\mathbf{28.101}$  & $18.874$   \\ 
			lap & $15.629$  & $15.843$  & $20.540$  & $\mathbf{30.216}$  & $22.953$   & ~ & $14.163$  & $14.366$  & $18.997 $ & $\mathbf{28.679}$  & $21.895$   & ~ & $13.040$  & $13.267$  & $17.702$  & $\mathbf{28.517}$  & $21.537$   \\ 
			nor lap & $14.898$  & $15.096$  & $19.548$  & $\mathbf{32.872}$  & $22.313$   & ~ & $13.623$  & $13.796$  & $18.150$  & $\mathbf{31.844}$  & $20.459$   & ~ & $12.612$  & $12.805$  & $16.996$  & $\mathbf{22.425}$  & $20.146$   \\ 
			row nor adj & $15.555$  & $15.780$  & $20.518$  & $\mathbf{30.081}$  & $21.160$   & ~ & $14.109$  & $14.308$  & $19.040$  & $\mathbf{29.733}$  & $20.066$   & ~ & $12.992 $ & $13.203$  & $17.857$  & $\mathbf{28.637}$  & $19.451$   \\ 
			sym nor adj & $14.920$  & $15.109$  & $19.537$  & $\mathbf{29.743}$  & $20.708$   & ~ & $13.641$  & $13.813$  & $18.142$  & $\mathbf{29.097}$  & $21.623$   & ~ & $12.629$  & $12.824$  & $16.996$  & $\mathbf{28.842}$  & $18.541$   \\ 
			\bottomrule[1.5pt]
		\end{tabular}
	}
\end{table*}

\indent \emph{2) Viedo denoising:} We conduct video denoising experiments using the publicly available REDS dataset \cite{son2021ntire}. Five representative videos (Viedo.~08, 09, 11, 27, and 29) are selected for testing. For each sequence, three consecutive frames are extracted to form one time-varying video signal. Each frame is first resized to $512 \times 512$ pixels and then divided into non-overlapping sub-images of size $16 \times 16$. Each sub-image is vectorized and concatenated across the three frames, resulting in a time-varying graph signal of dimension $256 \times 3$. We construct a 4-nearest neighbor (4-NN) graph. The filter coefficients are initialized to one, and all learnable parameters are initialized to $0.5$. The network is trained for $800$ epochs with a learning rate of $0.01$. To quantitatively evaluate denoising performance, we adopt three widely used image quality metrics: MSE, peak SNR (PSNR), and structural similarity index (SSIM). The PSNR is defined as
\begin{align}
	\mathrm{PSNR} = 10\mathrm{log}_{10}
	\left(\frac{255^2}{\rm MSE}\right).
\end{align}

\indent Table~\ref{tab:Gradient Descent denoise viedo} reports the MSE, PSNR, and SSIM results for the five video sequences under $\sigma =45 $. The results show that the proposed DMPJFRFT-based denoising schemes consistently outperform the compared methods across all test sequences, demonstrating their superior capability in recovering fine spatial and temporal details.

\begin{table*}[htbp]
	\centering
	\caption{Denoising performance of gradient descent-based methods on five representative videos (Viedo 08, 09, 11, 27, and 29) from the REDS dataset.
		\label{tab:Gradient Descent denoise viedo}}
	\resizebox{1.0\linewidth}{!}{
		\begin{tabular}{cccccccccccc}
			\toprule[1.5pt]
			\multicolumn{1}{c}{\multirow{2}{*}{Viedo 08}} & \multicolumn{3}{c}{Frame 1} && \multicolumn{3}{c}{Frame 2} && \multicolumn{3}{c}{Frame 3} \\
			\cmidrule{2-4} \cmidrule{6-8} \cmidrule{10-12}
			~ & MSE & PSNR & SSIM && MSE & PSNR & SSIM && MSE & PSNR & SSIM \\ \midrule
			2D GFRFT & $4.865\times 10^{1}$ & $31.260$  & $0.926807$   & ~ & $7.744\times 10^{1}$ & $29.241$  & $0.945602$   & ~ & $4.873\times 10^{1}$ & $31.253$  & $0.934619$   \\ 
			2D GBFRFT & $4.891\times 10^{1}$ & $31.237$  & $0.927338$   & ~ & $7.700\times 10^{1}$ & $29.266$  & $0.946263$   & ~ & $4.911\times 10^{1}$ & $31.219$  & $0.934901 $  \\ 
			JFRFT & $5.449\times 10^{-1}$ & $50.768$  & $0.998790$   & ~ & $1.848$ & 45.465  & $0.996414$   & ~ & $2.634$ & $43.925$  & $0.995053$   \\ 
			DMPJFRFT-I-I & $3.603\times 10^{-5}$ & $92.565$  & $\mathbf{1.000000}$   & ~ & $5.103\times 10^{-5}$ & $91.052$  & $\mathbf{1.000000}$   & ~ & $5.880\times 10^{-5}$ & $90.437$  & $\mathbf{1.000000}$   \\
			DMPJFRFT-I-II & $\mathbf{5.249\times 10^{-6}}$ & $\mathbf{100.930}$  & $\mathbf{1.000000}$   & ~ & $\mathbf{3.862\times 10^{-6}}$ & $\mathbf{102.263}$  & $\mathbf{1.000000}$   & ~ & $\mathbf{7.137\times 10^{-6}}$ & $\mathbf{99.596}$  & $\mathbf{1.000000}$   \\
			\bottomrule[1.5pt]
			\toprule[1.5pt]
			\multicolumn{1}{c}{\multirow{2}{*}{Viedo 09}} & \multicolumn{3}{c}{Frame 1} && \multicolumn{3}{c}{Frame 2} && \multicolumn{3}{c}{Frame 3} \\
			\cmidrule{2-4} \cmidrule{6-8} \cmidrule{10-12}
			~ & MSE & PSNR & SSIM && MSE & PSNR & SSIM && MSE & PSNR & SSIM \\ \midrule
			2D GFRFT & $3.460$ & $42.741$  & $0.986918$   & ~ & $4.831$ & $41.291$  & $0.993086$   & ~ & $3.549$ & $42.630$  & $0.986919$   \\ 
			2D GBFRFT & $4.127$ & $41.974$  & $0.986405$   & ~ & $4.848$ & $41.276$  & $0.993085$   & ~ & $4.141$ & $41.960$  & $0.986415$   \\ 
			JFRFT & $2.993\times 10^{-2}$ & $63.370$  & $0.999934$   & ~ & $5.282\times 10^{-2}$ & $60.903$  & $0.999880$   & ~ & $1.109\times 10^{-2}$ & $67.682$  & $0.999963$   \\ 
			DMPJFRFT-I-I & $1.013\times 10^{-5}$ & $98.075$  & $\mathbf{1.000000}$   & ~ & $7.616\times 10^{-6}$ & $99.314$  & $\mathbf{1.000000}$   & ~ & $8.122\times 10^{-6}$ & $99.034$  & $\mathbf{1.000000}$   \\
			DMPJFRFT-I-II & $\mathbf{1.170\times 10^{-6}}$ & $\mathbf{107.448}$  & $\mathbf{1.000000}$   & ~ & $\mathbf{5.982\times 10^{-7}}$ & $\mathbf{110.363} $ & $\mathbf{1.000000}$   & ~ & $\mathbf{9.282\times 10^{-7}}$ & $\mathbf{108.455}$  & $\mathbf{1.000000}$   \\ 
			\bottomrule[1.5pt]
			\toprule[1.5pt]
			\multicolumn{1}{c}{\multirow{2}{*}{Viedo 11}} & \multicolumn{3}{c}{Frame 1} && \multicolumn{3}{c}{Frame 2} && \multicolumn{3}{c}{Frame 3} \\
			\cmidrule{2-4} \cmidrule{6-8} \cmidrule{10-12}
			~ & MSE & PSNR & SSIM && MSE & PSNR & SSIM && MSE & PSNR & SSIM \\ \midrule
			2D GFRFT & $1.140\times 10^{2}$ & $27.560$  & $0.880162 $  & ~ & $1.334\times 10^{2}$ & $26.879$  & $0.916380$   & ~ & $1.141\times 10^{2}$ & $27.556$  & $0.886830$   \\ 
			2D GBFRFT & $1.127\times 10^{2}$ & $27.612$  & $0.882048$   & ~ & $1.318\times 10^{2}$ & $26.932$  & $0.917313$   & ~ & $1.128\times 10^{2}$ & $27.608$  & $0.888590$   \\ 
			JFRFT & $2.034$ & $45.047$  & $0.996583$   & ~ & $2.964$ & $43.412$  & $0.995574$   & ~ & $3.981$ & $42.131$  & $0.993599$   \\ 
			DMPJFRFT-I-I & $7.181\times 10^{-5}$ & $89.569$  & $\mathbf{1.000000}$   & ~ & $1.344\times 10^{-4}$ & $86.847$  & $0.999999$   & ~ & $1.753\times 10^{-4}$ & $85.694$  & $0.999999 $  \\ 
			DMPJFRFT-I-II & $\mathbf{4.109\times 10^{-6}}$ & $\mathbf{101.993}$  & $\mathbf{1.000000}$   & ~ & $\mathbf{5.547\times 10^{-6}}$ & $\mathbf{100.690}$  & $\mathbf{1.000000}$   & ~ & $\mathbf{1.015\times 10^{-5}}$ & $\mathbf{98.066}$  & $\mathbf{1.000000} $  \\ 
			\bottomrule[1.5pt]
			\toprule[1.5pt]
			\multicolumn{1}{c}{\multirow{2}{*}{Viedo 27}} & \multicolumn{3}{c}{Frame 1} && \multicolumn{3}{c}{Frame 2} && \multicolumn{3}{c}{Frame 3} \\
			\cmidrule{2-4} \cmidrule{6-8} \cmidrule{10-12}
			~ & MSE & PSNR & SSIM && MSE & PSNR & SSIM && MSE & PSNR & SSIM \\ \midrule
			2D GFRFT & $6.107\times 10^{1}$ & $30.273$  & $0.924801$   & ~ & $1.145\times 10^{2}$ & $27.543$  & $0.897960$   & ~ & $6.131\times 10^{1}$ & $30.255$  & $0.925035$   \\ 
			2D GBFRFT & $6.020\times 10^{1}$ & $30.334$  & $0.926680$   & ~ & $1.125\times 10^{2}$ & $27.619$  & $0.899891$   & ~ & $5.998\times 10^{1}$ & $30.350$  & $0.927568$   \\ 
			JFRFT & $8.758\times 10^{-1}$ & $48.707$  & $0.998545$   & ~ & $1.973$ & $45.181$  & $0.996993$   & ~ & $1.653$ & $45.949$  & $0.997599$   \\
			DMPJFRFT-I-I & $1.239\times 10^{-4}$ & $87.201$  & $0.999999$   & ~ & $2.597\times 10^{-4}$ & $83.986$  & $0.999999$   & ~ & $2.497\times 10^{-4}$ & $84.156$  & $0.999998$   \\ 
			DMPJFRFT-I-II & $\mathbf{3.789\times 10^{-6}}$ & $\mathbf{102.345}$  & $\mathbf{1.000000}$   & ~ & $\mathbf{1.610\times 10^{-5}}$ & $\mathbf{96.062}$  & $\mathbf{1.000000}$   & ~ & $\mathbf{3.931\times 10^{-6}}$ & $\mathbf{102.185}$  & $\mathbf{1.000000}$   \\ 
			\bottomrule[1.5pt]
			\toprule[1.5pt]
			\multicolumn{1}{c}{\multirow{2}{*}{Viedo 29}} & \multicolumn{3}{c}{Frame 1} && \multicolumn{3}{c}{Frame 2} && \multicolumn{3}{c}{Frame 3} \\
			\cmidrule{2-4} \cmidrule{6-8} \cmidrule{10-12}
			~ & MSE & PSNR & SSIM && MSE & PSNR & SSIM && MSE & PSNR & SSIM \\ \midrule
			2D GBFRFT & $9.939$ & $38.157$  & $0.960086$   & ~ & $1.778\times 10^{1}$ & $35.633$  & $0.961815$   & ~ & $9.976$ & $38.141$  & $0.962655$   \\ 
			2D GFRFT & $1.019\times 10^{1}$ & $38.047$  & $0.957525$   & ~ & $1.791\times 10^{1}$ & $35.600$  & $0.961386$   & ~ & $1.022\times 10^{1}$ & $38.035$  & $0.959982  $ \\ 
			JFRFT & $1.426\times 10^{-1}$ & $56.590$  & $0.999347 $  & ~ & $3.919\times 10^{-1}$ & $52.199$  & $0.998402$   & ~ & $3.515\times 10^{-1}$ & $52.672$  & $0.998605$   \\ 
			DMPJFRFT-I-I & $9.221\times 10^{-5}$ & $88.483$  & $0.999999$   & ~ & $8.088\times 10^{-5}$ & $89.052$  & $0.999999$   & ~ & $1.040\times 10^{-4}$ & $87.960$  & $0.999998$   \\ 
			DMPJFRFT-I-II & $\mathbf{8.535\times 10^{-7}}$ & $\mathbf{108.819}$  & $\mathbf{1.000000}$   & ~ & $\mathbf{8.750\times 10^{-7}}$ & $\mathbf{108.711}$  & $\mathbf{1.000000}$   & ~ & $\mathbf{2.428\times 10^{-6}}$ & $\mathbf{104.278} $ & $\mathbf{1.000000} $  \\ 
			\bottomrule[1.5pt]
		\end{tabular}
	}
\end{table*}

\begin{figure*}
	\centering
	\includegraphics[width=7in]{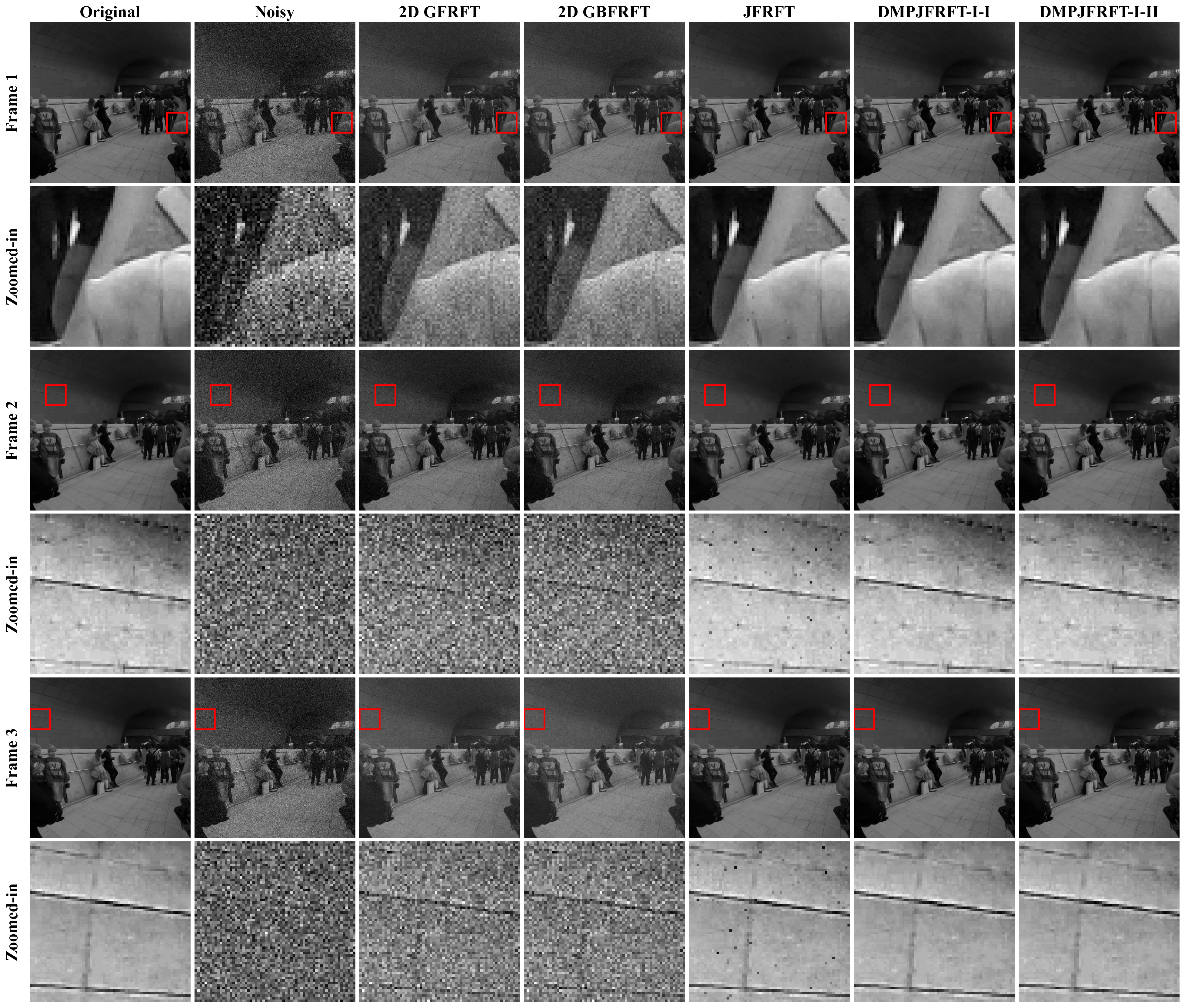}
	\caption{Visual comparison of denoising results for Video 29.}
	\label{fig:noisy_29}
\end{figure*}

\indent To further demonstrate the denoising performance, we provide visual comparisons. Fig.~\ref{fig:noisy_29} shows three consecutive frames from video 29, along with corresponding zoomed-in regions that highlight fine details. From both the full-frame views and the local magnified patches, it can be observed that the DMPJFRFT-based filtering method effectively removes almost all noise while preserving structural details and texture. In contrast, the other three compared methods leave noticeable residual noise points. 


\subsection{Debluring}
\indent The experimental setup for video deblurring is the same as \emph{2) Viedo denoising} that described in Section~\ref{Gradient Descent Denoising}, using the REDS dataset with paired sharp and blurred frames. Table~\ref{tab:Gradient Descent deblur viedo} reports the quantitative results for the selected video sequences, showing that the proposed DMPJFRFT-based method consistently outperforms the compared approaches in terms of MSE, PSNR and SSIM.

\begin{table*}[htbp]
	\centering
	\caption{Debluring performance of gradient descent-based methods on five representative videos (Viedo 08, 09, 11, 27, and 29) from the REDS dataset.
		\label{tab:Gradient Descent deblur viedo}}
	\resizebox{1.0\linewidth}{!}{
		\begin{tabular}{cccccccccccc}
			\toprule[1.5pt]
			\multicolumn{1}{c}{\multirow{2}{*}{Viedo 08}} & \multicolumn{3}{c}{Frame 1} && \multicolumn{3}{c}{Frame 2} && \multicolumn{3}{c}{Frame 3} \\
			\cmidrule{2-4} \cmidrule{6-8} \cmidrule{10-12}
			~ & MSE & PSNR & SSIM && MSE & PSNR & SSIM && MSE & PSNR & SSIM \\ \midrule
			2D GFRFT & $4.60\times10^{1}$ & $31.499$  & $0.934863$   & ~ & $7.45\times10^{1}$ & $29.406$  & $0.948618$   & ~ & $4.61\times10^{1}$ & $31.494$  & $0.943404$   \\
			2D GBFRFT & $4.62\times10^{1}$ & $31.489$  & $0.932986$   & ~ & $7.47\times10^{1}$ & $29.395$  & $0.948470$   & ~ & $4.61\times10^{1}$ & $31.492$  & $0.941310$   \\ 
			JFRFT & $2.28\times10^{-2}$ & $64.553$  & $0.999911$   & ~ & $7.36\times10^{-2}$ & $59.465$  & $0.999827$   & ~ & $1.31\times10^{-1}$ & $56.972$  & $0.999682$   \\
			DMPJFRFT-I-I & $4.55\times10^{-4}$ & $81.549$  & $0.999998$   & ~ & $8.12\times10^{-4}$ & $79.036$  & $0.999995$   & ~ & $7.85\times10^{-4}$ & $79.181$  & $0.999996$   \\ 
			DMPJFRFT-I-II & $\mathbf{1.23\times10^{-4}}$ & $\mathbf{87.218}$  & $\mathbf{0.999999}$   & ~ & $\mathbf{1.03\times10^{-4}}$ & $\mathbf{87.996}$  & $\mathbf{0.999999}$   & ~ & $\mathbf{9.46\times10^{-5}}$ & $\mathbf{88.371}$  & $\mathbf{0.999999}$   \\ 
			\bottomrule[1.5pt]
			\toprule[1.5pt]
			\multicolumn{1}{c}{\multirow{2}{*}{Viedo 09}} & \multicolumn{3}{c}{Frame 1} && \multicolumn{3}{c}{Frame 2} && \multicolumn{3}{c}{Frame 3} \\
			\cmidrule{2-4} \cmidrule{6-8} \cmidrule{10-12}
			~ & MSE & PSNR & SSIM && MSE & PSNR & SSIM && MSE & PSNR & SSIM \\ \midrule		
			2D GFRFT & $1.98$ & $45.163$  & $0.996352$   & ~ & $4.78$ & $41.333$  & $0.993171$   & ~ & $1.98$ & $45.163$  & $0.996344$   \\ 
			2D GBFRFT & $2.30$ & $44.521$  & $0.993869$   & ~ & $4.79$ & $41.324$  & $0.993164$   & ~ & $2.29$ & $44.533$  & $0.993918$   \\ 
			JFRFT & $6.49\times10^{-4}$ & $80.011 $ & $0.999998$   & ~ & $1.05\times10^{-3}$ & $77.932$  & $0.999997$   & ~ & $3.60\times10^{-4}$ & $82.570$  & $0.999998$   \\
			DMPJFRFT-I-I & $3.28\times10^{-4}$ & $82.972$  & $0.999999$   & ~ & $4.06\times10^{-4}$ & $82.043$  & $0.999998$   & ~ & $4.79\times10^{-4}$ & $81.324$  & $0.999998$   \\ 
			DMPJFRFT-I-II & $\mathbf{1.15\times10^{-4}}$ & $\mathbf{87.536}$  & $\mathbf{0.999999}$   & ~ & $\mathbf{5.57\times10^{-5}}$ & $\mathbf{90.671}$  & $\mathbf{1.000000}$   & ~ & $\mathbf{6.55\times10^{-5}}$ & $\mathbf{89.971}$  & $\mathbf{1.000000}$   \\ 
			\bottomrule[1.5pt]
			\toprule[1.5pt]
			\multicolumn{1}{c}{\multirow{2}{*}{Viedo 11}} & \multicolumn{3}{c}{Frame 1} && \multicolumn{3}{c}{Frame 2} && \multicolumn{3}{c}{Frame 3} \\
			\cmidrule{2-4} \cmidrule{6-8} \cmidrule{10-12}
			~ & MSE & PSNR & SSIM && MSE & PSNR & SSIM && MSE & PSNR & SSIM \\ \midrule
			2D GFRFT & $1.08\times10^{2}$ & $27.789$  & $0.891081$   & ~ & $1.27\times10^{2}$ & $27.083$  & $0.920668$   & ~ & $1.08\times10^{2}$ & $27.788$  & $0.897674$   \\ 
			2D GBFRFT & $1.08\times10^{2}$ & $27.806$  & $0.890285$   & ~ & $1.27\times10^{2}$ & $27.094$  & $0.920631$   & ~ & $1.08\times10^{2}$ & $27.807$  & $0.896928$   \\ 
			JFRFT & $1.01\times10^{-1}$ & $58.081$  & $0.999861$   & ~ & $1.48\times10^{-1}$ & $56.435$  & $0.999734$   & ~ & $1.91\times10^{-1}$ & $55.310$  & $0.999635$   \\ 
			DMPJFRFT-I-I & $4.52\times10^{-4}$ & $81.581$  & $0.999998$   & ~ & $7.05\times10^{-4}$ & $79.647$  & $0.999995$   & ~ & $6.40\times10^{-4}$ & $80.070$  & $0.999997$   \\ 
			DMPJFRFT-I-II & $\mathbf{9.74\times10^{-5}}$ & $\mathbf{88.247}$  & $\mathbf{0.999999}$   & ~ & $\mathbf{7.83\times10^{-5}}$ & $\mathbf{89.190}$  & $\mathbf{0.999999}$   & ~ & $\mathbf{8.28\times10^{-5}}$ & $\mathbf{88.950}$  & $\mathbf{0.999999}$   \\
			\bottomrule[1.5pt]
			\toprule[1.5pt]
			\multicolumn{1}{c}{\multirow{2}{*}{Viedo 27}} & \multicolumn{3}{c}{Frame 1} && \multicolumn{3}{c}{Frame 2} && \multicolumn{3}{c}{Frame 3} \\
			\cmidrule{2-4} \cmidrule{6-8} \cmidrule{10-12}
			~ & MSE & PSNR & SSIM && MSE & PSNR & SSIM && MSE & PSNR & SSIM \\ \midrule
			2D GFRFT & $5.74\times10^{1}$ & $30.538$  & $0.932755$   & ~ & $1.08\times10^{2}$ & $27.788$  & $0.904006$   & ~ & $5.74\times10^{1}$ & $30.540$  & $0.933115$   \\
			2D GBFRFT & $5.73\times10^{1}$ & $30.551$  & $0.931649$   & ~ & $1.08\times10^{2}$ & $27.786$  & $0.904105$   & ~ & $5.73\times10^{1}$ & $30.551$  & $0.932029$   \\ 
			JFRFT & $2.96\times10^{-2}$ & $63.415$  & $0.999935$   & ~ & $6.08\times10^{-2}$ & $60.292$  & $0.999877$   & ~ & $9.00\times10^{-2}$ & $58.588$  & $0.999882$   \\ 
			DMPJFRFT-I-I & $3.42\times10^{-4}$ & $82.792$  & $0.999998 $  & ~ & $6.38\times10^{-4}$ & $80.081$  & $0.999996$   & ~ & $6.17\times10^{-4}$ & $80.228$  & $0.999997$   \\ 
			DMPJFRFT-I-II & $\mathbf{5.51\times10^{-5}}$ & $\mathbf{90.719}$  & $\mathbf{1.000000}$   & ~ & $\mathbf{3.50\times10^{-5}}$ & $\mathbf{92.686}$  & $\mathbf{1.000000}$   & ~ & $\mathbf{3.77\times10^{-5}}$ & $\mathbf{92.373}$  & $\mathbf{1.000000}$   \\ 
			\bottomrule[1.5pt]
			\toprule[1.5pt]
			\multicolumn{1}{c}{\multirow{2}{*}{Viedo 29}} & \multicolumn{3}{c}{Frame 1} && \multicolumn{3}{c}{Frame 2} && \multicolumn{3}{c}{Frame 3} \\
			\cmidrule{2-4} \cmidrule{6-8} \cmidrule{10-12}
			~ & MSE & PSNR & SSIM && MSE & PSNR & SSIM && MSE & PSNR & SSIM \\ \midrule
			2D GFRFT & $9.24$ & $38.474$  & $0.965326$   & ~ & $1.72\times10^{1}$ & $35.785$  & $0.963183$   & ~ & $9.24$ & $38.473$  & $0.967920$   \\ 
			2D GBFRFT & $9.38$ & $38.407$  & $0.963393$   & ~ & $1.71\times10^{1}$ & $35.792$  & $0.963196$   & ~ & $9.39$ & $38.406$  & $0.966096$   \\ 
			JFRFT & $5.57\times10^{-3}$ & $70.676$  & $0.999963$   & ~ & $1.96\times10^{-2}$ & $65.214$  & $0.999907$   & ~ & $1.80\times10^{-2}$ & $65.586$  & $0.999927$   \\ 
			DMPJFRFT-I-I & $1.98\times10^{-4}$ & $85.161$  & $0.999998$   & ~ & $3.70\times10^{-4}$ & $82.453$  & $0.999996$   & ~ & $3.53\times10^{-4}$ & $82.647$  & $0.999997$   \\ 
			DMPJFRFT-I-II & $\mathbf{4.58\times10^{-5}}$ & $\mathbf{91.522}$  & $\mathbf{1.000000}$   & ~ & $\mathbf{2.50\times10^{-5}}$ & $\mathbf{94.151}$  & $\mathbf{1.000000}$   & ~ & $\mathbf{2.27\times10^{-5}}$ & $\mathbf{94.564}$  & $\mathbf{1.000000}$   \\ 
			\bottomrule[1.5pt]
		\end{tabular}
	}
\end{table*}

\begin{figure*}
	\centering
	\includegraphics[width=7in]{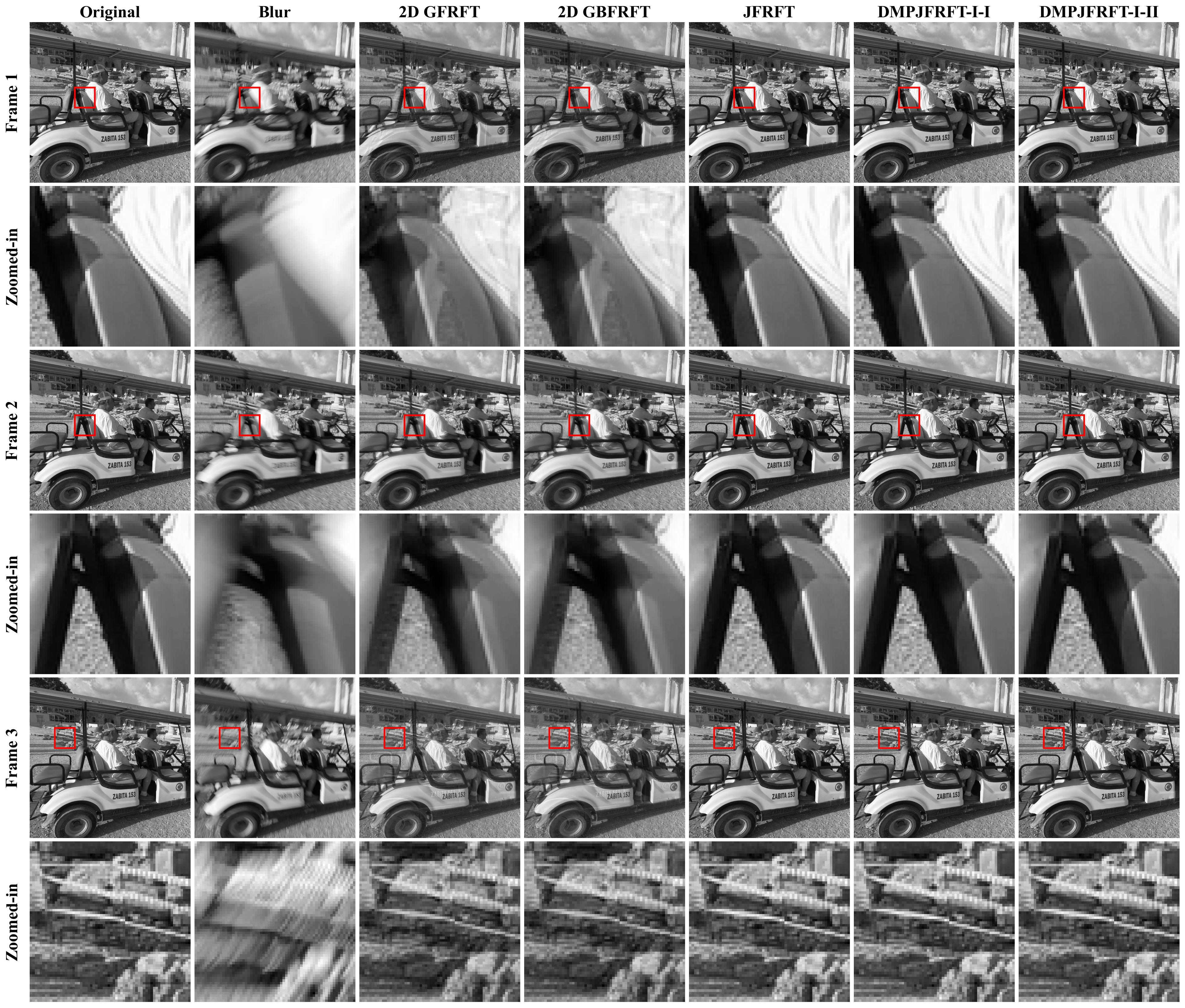}
	\caption{Visual comparison of debluring results for Video 11.}
	\label{fig:blur_11}
\end{figure*}

\indent To further illustrate the restoration quality, Fig.~\ref{fig:blur_11} presents the deblurring results for video~11. It can be observed that the proposed DMPJFRFT-based methods (DMPJFRFT-I-I and DMPJFRFT-I-II) and the JFRFT-based method both achieve significantly better visual quality than the 2D GFRFT and 2D GBFRFT, effectively removing blur while preserving edge sharpness and structural details. Notably, the visual results of JFRFT and the two DMPJFRFT variants appear quite similar. This is mainly because, under the current experimental setting where complete prior information of the signal is available, both methods can reconstruct the spatio-temporal frequency structures with high fidelity. However, quantitative evaluation metrics in Table~\ref{tab:Gradient Descent deblur viedo} still indicate that our proposed methods perform slightly better, suggesting their stronger adaptability.

\begin{figure}
	\centering
	\includegraphics[width=3.5in]{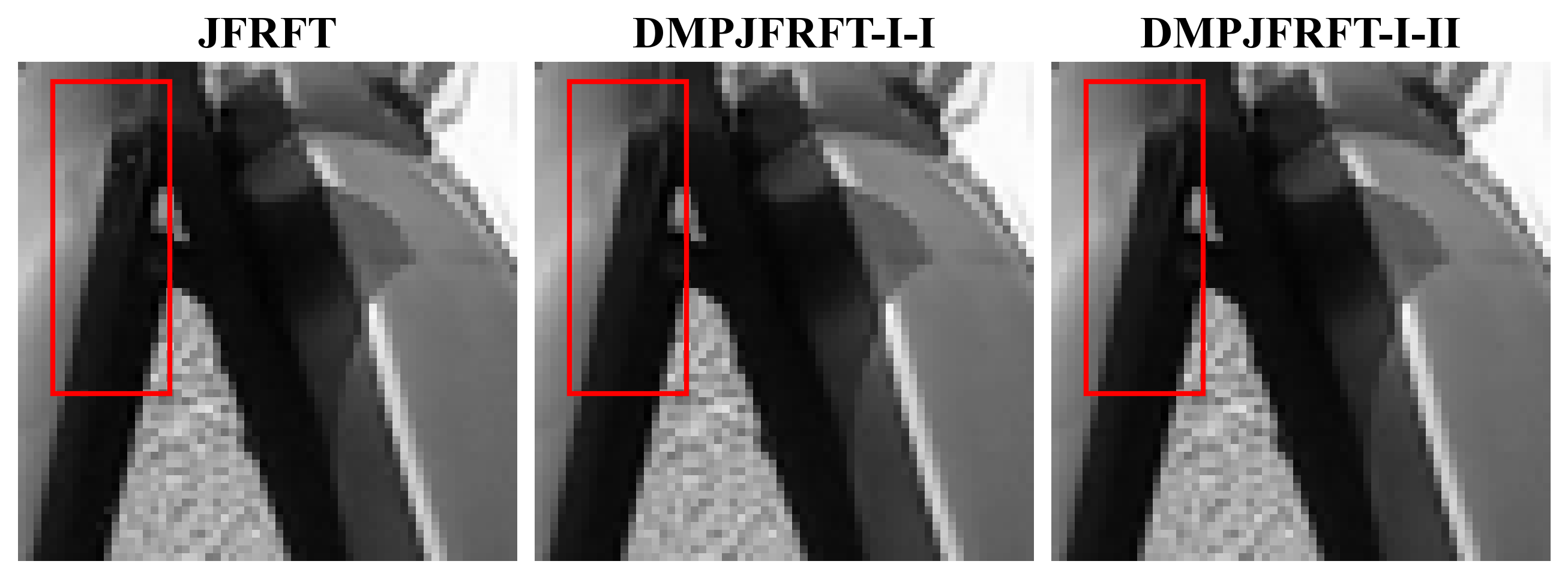}
	\caption{Zoomed-in comparison of local details in the second frame among JFRFT, DMPJFRFT-I-I, and DMPJFRFT-I-II for Video 08.}
	\label{fig:blur_11_zoomed}
\end{figure}

\indent To provide a more detailed comparison, Fig.~\ref{fig:blur_11_zoomed} presents the zoomed-in regions of the second frame, where red rectangles highlight areas with noticeable local details. Although the overall visual appearance of JFRFT, DMPJFRFT-I-I, and DMPJFRFT-I-II is similar, the proposed methods still produce slightly finer texture restoration within these highlighted regions.

\section{Neural Network-Based DMPJFRFT Filtering}   \label{sec5}
\indent Although the gradient descent-based DMPJFRFT filtering scheme demonstrates strong denoising and deblurring capabilities, it still faces certain limitations in practical applications. Specifically, this method requires complete prior knowledge of the clean signal, which is often unavailable in real-world scenarios. To address this limitation, we propose a neural network-based framework, termed DMPJFRFTNet, which learns to perform filtering with only partial prior information. 

\begin{figure}
	\centering
	\includegraphics[width=3.5in]{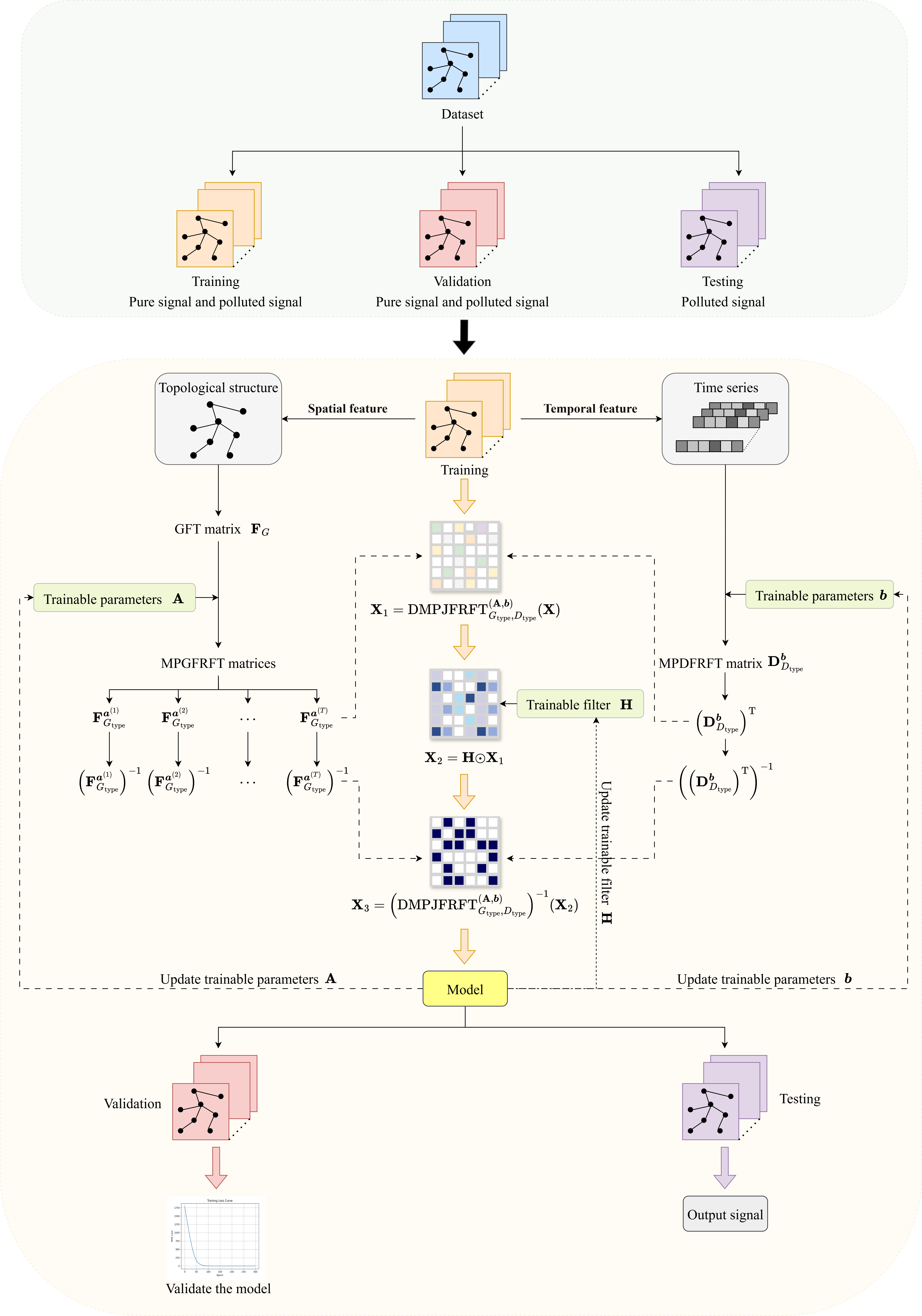}
	\caption{Overall architecture of the proposed DMPJFRFTNet.}
	\label{fig:DMPJFRFTNet}
\end{figure}

\indent In the proposed DMPJFRFTNet, the time-varying graph signals are divided into training, validation, and testing sets. The training and validation sets contain paired clean and corrupted signals, which are used to optimize the learnable parameters of the network, including the fractional orders and filter coefficients. During testing, however, only corrupted signals are provided, allowing the model to generalize and infer the underlying clean signal without explicit prior knowledge. The overall architecture of the proposed DMPJFRFTNet are illustrated in Fig.~\ref{fig:DMPJFRFTNet}.

\indent The overall architecture of DMPJFRFTNet integrates the DMPJFRFT as a differentiable fully connected layer within the network. A time-varying graph signal is represented as a matrix $\mathbf{X} \in \mathbb{C}^{N \times TM}$, where $N$ denotes the number of vertices, $T$ represents the number of time frames for each sample, and $M$ is the number of signal samples in a batch. Equivalently, $\mathbf{X}$ can be viewed as $M$ time-varying graph signals ${\mathbf{X}_1, \mathbf{X}_2, \ldots, \mathbf{X}_M}$, each of dimension $N \times T$. 

\indent Within this framework, the DMPJFRFT transformation module performs a joint spectral representation of each time-varying graph signal. The forward DMPJFRFT layer, implemented as a fully connected operation, transforms $\mathbf{X}$ into its spectral-domain representation 
$\mathbf{Y} = \mathbf{F}_{J,G_{\mathrm{type}},D_{\mathrm{type}}}^{(\mathbf{A}, \boldsymbol{b})}(\mathbf{X}) \in \mathbb{C}^{N \times TM}$, 
where $(\mathbf{A}, \boldsymbol{b})$ are the learnable fractional orders in the graph and temporal dimensions. These parameters are optimized during training, allowing the layer to adaptively adjust the spectral representation to different signal characteristics. In this domain, a learnable diagonal filter $\mathbf{H}$ enhances meaningful signal components while suppressing noise or blur. Finally, the inverse DMPJFRFT reconstructs the filtered signal back to the vertex domain, yielding the restored time-varying graph signal $\widetilde{\mathbf{X}}$. The detailed training and inference procedure of DMPJFRFTNet is summarized in Algorithm~\ref{alg:dmpjfrftnet_algo}.

\begin{algorithm}[!htbp]
	\caption{Training and Testing of DMPJFRFTNet}
	\label{alg:dmpjfrftnet_algo}
	\begin{algorithmic}[1]
		\REQUIRE Time-varying graph signal $\mathbf{X}\in\mathbb{C}^{N\times TM}$, noisy observation $\mathbf{X}_{\mathrm{noisy}}$, learning rate $\gamma$, training epochs $E$, group size $G$
		\ENSURE Learned transform parameters $(\mathbf{A}^*,\boldsymbol{b}^*)$, learned spectral filter $\mathbf{H}^*$, restored signal $\widetilde{\mathbf{X}}$, evaluation metrics
		
		\STATE Split dataset into training, validation, and test sets
		\STATE Initialize fractional orders $\mathbf{A}\leftarrow\mathbf{A}^{(0)}$, $\boldsymbol{b}\leftarrow\boldsymbol{b}^{(0)}$ and filter $\mathbf{H}\leftarrow\mathbf{H}^{(0)}$
		
		\FOR{epoch = 1 to $E$}
		\STATE \textbf{Training Phase:}
		\FOR{each training batch}
		\STATE Forward pass: \\
		 $\widetilde{\mathbf{X}}_{\mathrm{batch}} \leftarrow \text{DMPJFRFTNet}(\mathbf{X}_{\mathrm{noisy,batch}}, \mathbf{A}, \boldsymbol{b}, \mathbf{H})$
		\STATE Compute training loss: \\ $\mathcal{L}_{\mathrm{train}} \leftarrow \|\widetilde{\mathbf{X}}_{\mathrm{batch}} - \mathbf{X}_{\mathrm{batch}}\|_F^2 / NT$
		\STATE Backpropagation: \\
		update $(\mathbf{A}, \boldsymbol{b}, \mathbf{H})$ using Adam optimizer
		\ENDFOR
		
		\STATE \textbf{Validation Phase:}
		\STATE Forward pass on validation set
		\STATE Compute validation metrics: (MSE, SNR) for real-data and (MSE, PSNR, SSIM) for videos
		\ENDFOR
		
		\STATE \textbf{Testing:}
		\STATE Load best parameters $(\mathbf{A}^*, \boldsymbol{b}^*, \mathbf{H}^*)$
		\STATE Forward pass on test set: \\
		$\widetilde{\mathbf{X}} \leftarrow \text{DMPJFRFTNet}(\mathbf{X}_{\mathrm{noisy,test}}, \mathbf{A}^*, \boldsymbol{b}^*, \mathbf{H}^*)$
		\STATE Compute evaluation metrics: (MSE, SNR) for real-data and (MSE, PSNR, SSIM) for videos
	\end{algorithmic}
\end{algorithm}

\subsection{Denoising}  \label{Neural Network Denoising}   
\indent \emph{1) Real-data denoising:} We conduct denoising experiments on real-world time-varying graph signals using the same datasets described in Section~\ref{Gradient Descent Denoising}. For each dataset, we truncate the first $1500$ time series, and treat every $6$ consecutive columns as a single time-varying graph signal. From the resulting set, $20\%$ of signals are randomly selected as the test set, while the remaining $80\%$ are split into training and validation sets with a ratio of $8:2$. The learning rate is set to $0.001$. For the SST dataset, the number of training epochs is set to $100$, whereas for the other three datasets, $200$ epochs are used. The initial values of the spectral filter coefficients are set to $1$, and the initial fractional orders in both graph and temporal dimensions are set to $0.5$.  

\indent We compare the denoising performance of 2D GFRFTNet, 2D GBFRFTNet, and JFRFTNet under five types of GSOs and different noise levels. The results are summarized in Table~\ref{tab:Neural Network denoise signal}. It can be observed that the proposed DMPJFRFT-based neural network consistently achieves the highest SNR across different GSOs and noise levels.

\begin{table}[htbp]
	\caption{SNR performance of neural network-based denoising methods (2D GFRFTNet, 2D GBFRFTNet, and JFRFTNet) under various GSOs for real-world graph signals.
		\label{tab:Neural Network denoise signal}}
	\centering
	\resizebox{1.0\linewidth}{!}{
		\begin{tabular}{cccccc}
		\toprule[1.5pt]
		~ & 2D GFRFTNet & 2D GBFRFTNet & JFRFTNet & DMPJFRFT-I-INet & DMPJFRFT-I-IINet \\ \midrule
		\multicolumn{1}{c}{\multirow{1}{*}{PEMSD7(M)}} & \multicolumn{5}{c}{$\sigma=40$ (SNR=$3.805$)} \\
		\cmidrule{2-6}
			adj & $16.090$ & $16.102$ & $17.345$ & $\mathbf{17.490}$ & $16.695$  \\ 
			lap & $16.111$ & $16.123$ & $17.384$ & $\mathbf{17.466}$ & $16.719$  \\ 
			nor lap & $16.073$ & $16.086$ & $17.303$ & $\mathbf{17.419}$ & $16.643$  \\ 
			row nor adj & $16.065$ & $16.077$ & $17.289$ & $\mathbf{17.417}$ & $16.676$  \\ 
			sym nor adj & $16.065$ & $16.077$ & $17.289$ & $\mathbf{17.417}$ & $16.676$   \\ \midrule
			~ & \multicolumn{5}{c}{$\sigma=50$ (SNR=$1.866$)} \\
			\cmidrule{2-6}
			adj & $15.813$ & $15.826$ & $16.882$ & $\mathbf{17.022}$ & $16.421$  \\ 
			lap & $15.827$ & $15.839$ & $16.911$ & $\mathbf{17.002}$ & $16.461$  \\ 
			nor lap & $15.799$ & $15.813$ & $16.835$ & $\mathbf{16.968}$ & $16.401$  \\ 
			row nor adj & $15.795$ & $15.808$ & $16.829$ & $\mathbf{16.957}$ & $16.408$  \\ 
			sym nor adj & $15.795$ & $15.808$ & $16.829$ & $\mathbf{16.957}$ & $16.408$   \\  \midrule
			~ & \multicolumn{5}{c}{$\sigma=60$ (SNR=$0.283$)} \\
			\cmidrule{2-6}
			adj & $15.568$ & $15.561$ & $16.578$ & $\mathbf{16.665}$ & $16.18$  \\ 
			lap & $15.577$ & $15.583$ & $16.601$ & $\mathbf{16.679}$ & $16.214$  \\ 
			nor lap & $15.558$ & $15.552$ & $16.531$ & $\mathbf{16.631}$ & $16.159$  \\ 
			row nor adj & $15.555$ & $15.546$ & $16.529$ & $\mathbf{16.648}$ & $16.173$  \\ 
			sym nor adj & $15.555$ & $15.546$ & $16.529$ & $\mathbf{16.648}$ & $16.173$   \\ 
		\bottomrule[1.5pt]
		\toprule[1.5pt]
		\multicolumn{1}{c}{\multirow{1}{*}{PEMS08}} & \multicolumn{5}{c}{$\sigma=200$ (SNR=$1.741$)} \\
		\cmidrule{2-6}
			adj & $6.889$ & $6.897$ & $11.356$ & $\mathbf{12.382}$ & $11.388$  \\ 
			lap & $9.317$ & $9.32$ & $12.308$ & $13.95$ & $\mathbf{14.123}$  \\ 
			nor lap & $8.683$ & $8.686$ & $12.092$ & $13.539$ & $\mathbf{13.929}$  \\ 
			row nor adj & $9.246$ & $9.254$ & $12.297$ & $\mathbf{13.451}$ & $12.141$  \\ 
			sym nor adj & $8.694$ & $8.697$ & $12.098$ & $\mathbf{13.564}$ & $12.433$  \\ 
			  \midrule
			~ & \multicolumn{5}{c}{$\sigma=240$ (SNR=$0.157$)} \\
			\cmidrule{2-6}
			adj & $6.065$ & $6.073$ & $10.316$ & $\mathbf{11.790}$ & $10.430$  \\ 
			lap & $8.848$ & $8.851$ & $11.51$ & $\mathbf{13.471}$ & $12.155$  \\ 
			nor lap & $8.123$ & $8.126$ & $11.232$ & $\mathbf{12.972}$ & $11.919$  \\ 
			row nor adj & $8.758$ & $8.768$ & $11.496$ & $12.828$ & $\mathbf{12.927}$ \\ 
			sym nor adj & $8.135$ & $8.137$ & $11.243$ & $\mathbf{12.992}$ & $12.620$  \\  \midrule
			~ & \multicolumn{5}{c}{$\sigma=280$ (SNR=$-1.182$)} \\
			\cmidrule{2-6} 
			adj & $5.412$ & $5.419$ & $9.466$ & $\mathbf{11.132}$ & $9.049$  \\ 
			lap & $8.498$ & $8.5$ & $10.892$ & $\mathbf{13.098}$ & $11.768$  \\ 
			nor lap & $7.701$ & $7.704$ & $10.556$ & $\mathbf{12.518}$ & $11.301$  \\ 
			row nor adj & $8.389$ & $8.399$ & $10.872$ & $\mathbf{12.384}$ & $12.313$  \\ 
			sym nor adj & $7.712$ & $7.715$ & $10.570$ & $\mathbf{12.339}$ & $10.783$  \\ 
		\bottomrule[1.5pt]
		\toprule[1.5pt]
		\multicolumn{1}{c}{\multirow{1}{*}{Quality}} & \multicolumn{5}{c}{$\sigma=50$ (SNR=$3.148$)} \\
		\cmidrule{2-6}
			adj & $6.578$ & $6.576$ & $10.788$ & $\mathbf{10.859}$ & $8.415$  \\ 
			lap & $6.862$ & $6.867$ & $11.228$ & $\mathbf{12.015}$ & $11.189$  \\ 
			nor lap & $6.493$ & $6.53$ & $10.809$ & $\mathbf{11.977}$ & $10.16$  \\ 
			row nor adj & $7.322$ & $7.351$ & $11.207$ & $\mathbf{11.881}$ & $11.559$ \\ 
			sym nor adj & $6.020$ & $6.082$ & $10.495$ & $\mathbf{10.721}$ & $10.618$  \\ \midrule
		~ & \multicolumn{5}{c}{$\sigma=80$ (SNR=$-0.935$)} \\
		\cmidrule{2-6} 
			adj & $4.323$ & $4.320$ & $8.154$ & $\mathbf{8.318}$ & $7.525$  \\ 
			lap & $4.274$ & $4.273$ & $8.798$ & $\mathbf{9.746}$ & $9.336$  \\ 
			nor lap & $4.129$ & $4.168$ & $8.265$ & $\mathbf{9.677}$ & $8.678$  \\ 
			row nor adj & $4.869$ & $4.901$ & $8.803$ & $\mathbf{9.597}$ & $9.127$  \\ 
			sym nor adj & $3.705$ & $3.757$ & $7.65$ & $7.749$ & $\mathbf{8.302}$  \\ \midrule
		~ & \multicolumn{5}{c}{$\sigma=100$ (SNR=$-2.873$)} \\
		\cmidrule{2-6} 
			adj & $3.441$ & $3.438$ & $6.956$ & $\mathbf{7.078}$ & $6.871$  \\ 
			lap & $3.246$ & $3.244$ & $7.59$ & $\mathbf{8.544}$ & $8.393$  \\ 
			nor lap & $3.218$ & $3.259$ & $7.07$ & $\mathbf{8.553}$ & $7.962$  \\ 
			row nor adj & $3.854$ & $3.889$ & $7.647$ & $\mathbf{8.439}$ & $7.97$  \\ 
			sym nor adj & $2.841$ & $2.887$ & $6.378$ & $6.386$ & $\mathbf{7.234}$  \\
		\bottomrule[1.5pt]
		\toprule[1.5pt]
		\multicolumn{1}{c}{\multirow{1}{*}{SST}} & \multicolumn{5}{c}{$\sigma=15$ (SNR=$2.956$)} \\
		\cmidrule{2-6}
			adj & $11.807$ & $11.801$ & $15.846$ & $\mathbf{19.005}$ & $16.472$  \\ 
			lap & $15.015$ & $15.019$ & $19.145$ & $\mathbf{21.642}$ & $16.912$  \\ 
			nor lap & $14.210$ & $14.213$ & $17.670$ & $\mathbf{21.650}$ & $18.869$  \\ 
			row nor adj & $14.834$ & $14.836$ & $18.913$ & $\mathbf{21.543}$ & $19.836$  \\ 
			sym nor adj & $14.237$ & $14.235$ & $17.688$ & $\mathbf{20.887}$ & $18.891$  \\ \midrule
		~ & \multicolumn{5}{c}{$\sigma=20$ (SNR=$0.457$)} \\
		\cmidrule{2-6}
			adj & $10.442$ & $10.436$ & $14.464$ & $\mathbf{17.192}$ & $14.515$  \\ 
			lap & $13.543$ & $13.544$ & $17.778$ & $\mathbf{20.093}$ & $15.972$  \\ 
			nor lap & $12.936$ & $12.936$ & $16.515$ & $\mathbf{19.905}$ & $16.606$  \\ 
			row nor adj & $13.381$ & $13.381$ & $17.574$ & $\mathbf{19.822}$ & $17.907$  \\ 
			sym nor adj & $12.959$ & $12.957$ & $16.543$ & $\mathbf{19.174}$ & $16.631$  \\  \midrule
		~ & \multicolumn{5}{c}{$\sigma=25$ (SNR=$-1.481$)} \\
		\cmidrule{2-6}
			adj & $9.385$ & $9.378$ & $13.423$ & $\mathbf{15.713}$ & $12.995$  \\ 
			lap & $12.406$ & $12.404$ & $16.709$ & $\mathbf{18.772}$ & $14.869$  \\ 
			nor lap & $11.926$ & $11.923$ & $15.625$ & $\mathbf{18.332}$ & $14.437$  \\ 
			row nor adj & $12.257$ & $12.255$ & $16.523$ & $\mathbf{18.368}$ & $16.110$  \\ 
			sym nor adj & $11.945$ & $11.943$ & $15.659$ & $\mathbf{17.628}$ & $14.428$  \\
		\bottomrule[1.5pt]
		\end{tabular}
	}
\end{table}

\indent To further validate the effectiveness of the proposed DMPJFRFT-based models, we select the best results from the five GSOs for each method and compare them with several representative graph neural network approaches, including APPNP \cite{gasteiger2018predict}, ARMAconv \cite{gasteiger2018predict}, BernNet \cite{he2021bernnet}, ChebyNet \cite{Bianchi22Graph}, GAT \cite{velickovic2017graph}, GCN \cite{jiang2019semi}, LanczosNet \cite{liao2019lanczosnet}, Specformer \cite{Bo23Specformer}, SpectralCNN \cite{wang2022powerful,wang2022equivariant}, and UniMP \cite{shi2020masked}. The comparison methods are trained with a batch size of $16$, and their learning rate and number of epochs are kept consistent with the settings described above to ensure fairness. The denoising results are summarized in Table~\ref{tab:all Neural Network denoise signal}. Except for the Quality dataset, where LanczosNet achieves slightly better denoising performance, the proposed DMPJFRFT-based networks consistently outperform all baseline methods across the remaining datasets. 

\begin{table*}[htbp]
	\caption{Comparison of denoising performance between the proposed DMPJFRFTNet models (under optimal GSOs) and several state-of-the-art methods for graph signal denoising.
		\label{tab:all Neural Network denoise signal}}
	\centering
	\resizebox{1.0\linewidth}{!}{
		\begin{tabular}{ccccccccccccccccc}
			\toprule[1.5pt]
		~ & ~ & 2D GFRFTNet & 2D GBFRFTNet & JFRFTNet & APPNP & ARMAconv & BernNet & ChebyNet & GAT & GCN & LanczosNet & Specformer & SpectralCNN  & UniMP & DMPJFRFT-I-INet & DMPJFRFT-I-IINet \\ \midrule
		\multicolumn{1}{c}{\multirow{3}{*}{PEMSD7(M)}} & $\sigma=40$ & $16.111$ & $16.123$ & $17.384$  & $15.309$ & $13.455$ & $13.333$ & $14.343$ & $14.968$ & $14.944$ & $16.555$ & $14.817$ & $15.153$  & $15.580$  &  $\mathbf{17.490}$ & $16.719$  \\ 
		~ & $\sigma=50$ & $15.827$ & $15.839$ & $16.911$  & $14.454$ & $12.373$ & $12.234$ & $13.514$ & $14.461$ & $14.218$ & $15.900$ & $14.039$ & $14.429$  & $14.457$  & $\mathbf{17.022}$  & $16.461$  \\ 
		~ & $\sigma=60$ & $15.577$ & $15.583$ & $16.601$  & $14.254$ & $11.641$ & $11.492$ & $12.799$ & $14.087$ & $13.822$ & $15.406$ & $14.039$ & $13.888$  & $13.595$  & $\mathbf{16.679}$ & $16.214$  \\ \midrule
		\multicolumn{1}{c}{\multirow{3}{*}{PEMS08}} & $\sigma=200$ & $9.317$ & $9.320$ & $12.308$  & $10.701$ & $10.35$ & $10.682$ & $10.553$ & $10.331$ & $10.28$ & $12.945$ & $10.094$ & $12.434$  & $10.739$  & $13.95$ & $\mathbf{14.123}$  \\ 
		~ & $\sigma=240$ & $8.848$ & $8.851$ & $11.51$  & $9.609$ & $9.623$ & $9.595$ & $9.425$ & $9.364$ & $9.306$ & $12.565$ & $9.000$ & $11.886$  & $9.653$  & $\mathbf{13.471}$  & $12.927$  \\ 
		~ & $\sigma=280$ & $8.498$ & $8.500$ & $10.892$  & $8.763$ & $8.772$ & $8.753$ & $8.553$ & $8.578$ & $8.524$ & $12.209$ & $8.311$ & $11.426$  & $8.783$  & $\mathbf{13.098}$  & $12.313$  \\ \midrule
		\multicolumn{1}{c}{\multirow{3}{*}{Quality}} & $\sigma=50$ & $7.322$ & $7.351$ & $11.228$  & $10.095$ & $10.727$ & $10.685$ & $10.208$ & $9.923$ & $10.197$ & $\mathbf{12.337}$ & $10.746$ & $10.663$  & $9.878$  & $12.015$ & $11.559$  \\ 
		~ & $\sigma=80$ & $4.869$ & $4.901$ & $8.803$  & $7.296$ & $8.118$ & $7.981$ & $7.490$ & $6.711$ & $7.728$ & $\mathbf{10.197}$ & $8.279$ & $8.850$  & $6.832$  & $9.746$ & $9.336$  \\ 
		~ & $\sigma=100$ & $3.854$ & $3.889$ & $7.647$  & $5.837$ & $6.912$ & $6.775$ & $6.369$ & $5.605$ & $6.620$ & $\mathbf{9.162}$ & $7.275$ & $7.886$  & $5.609$  & $8.553$ & $8.393$  \\  \midrule
		\multicolumn{1}{c}{\multirow{3}{*}{SST}} & $\sigma=15$ & $15.015$ & $15.019$ & $19.145$  & $12.406$ & $12.014$ & $12.002$ & $12.112$ & $12.756$ & $12.366$ & $19.887$ & $13.014$ & $17.994$  & $12.536$  & $\mathbf{21.650}$ & $19.836$  \\ 
		~ & $\sigma=20$ & $13.543$ & $13.544$ & $17.778$  & $10.801$ & $10.373$ & $10.356$ & $10.518$ & $11.247$ & $10.827$ & $19.517$ & $12.310$ & $16.242$  & $11.137$  & $\mathbf{20.093}$ & $17.907$  \\ 
		~ & $\sigma=25$ & $12.406$ & $12.404$ & $16.709$  & $9.994$ & $9.343$ & $9.318$ & $9.476$ & $10.251$ & $9.777$ & $18.147$ & $12.050$ & $15.122$  & $10.176$  & $\mathbf{18.772}$ & $16.110$  \\ 
			\bottomrule[1.5pt]
		\end{tabular}
	}
\end{table*}

\indent \emph{2) Video denoising:} The video denoising experiments are conducted using five sequences from the publicly available GoPro dataset \cite{nah2017deep}, namely GOPR0384\_11\_02, GOPR0384\_11\_03, GOPR0869\_11\_00, GOPR0396\_11\_00, and GOPR0385\_11\_01. Each sequence contains 1100 frames, and every consecutive five frames are treated as a short video clip. Each frame is first resized to $128\times128$, then divided into $64$ non-overlapping $16\times16$ patches, each of which is vectorized and represented as a graph signal. A 4-NN graph is constructed for each patch. The noisy videos are generated by adding Gaussian noise with $\sigma = 45$. The learning rate is set to $0.001$, and the number of training epochs is $500$. We randomly select $20\%$ of the video clips for testing, while the remaining $80\%$ are divided into training and validation sets with a ratio of $8:2$. Table~\ref{tab:Neural Network denoise viedo} summarizes the average denoising performance of the proposed DMPJFRFTNet along with several representative baseline methods on the test set, evaluated using three standard image quality metrics: MSE, PSNR, and SSIM. The results indicate that DMPJFRFTNet attains the best overall performance, verifying the effectiveness of the proposed multiple-parameter joint fractional transform framework.

\begin{table*}[htbp]
	\caption{Denoising performance of neural network-based methods on five representative videos from the GoPro dataset.
		\label{tab:Neural Network denoise viedo}}
	\centering
	\resizebox{1.0\linewidth}{!}{
	\begin{tabular}{ccccccccccccccccc}
		\toprule[1.5pt]
		~ & ~ & 2D GFRFTNet & 2D GBFRFTNet & JFRFTNet & APPNP & ARMAconv & BernNet & ChebyNet & GAT & GCN & LanczosNet & Specformer & SpectralCNN & UniMP & DMPJFRFT-I-INet & DMPJFRFT-I-IINet \\ \midrule
		\multicolumn{1}{c}{\multirow{3}{*}{GOPR0384\_11\_02}} &
		MSE & $566.050$  & $160.966$  & $97.420$  & $161.428$  & $167.422$  & $163.181$  & $169.510$  & $159.485$  & $165.639$  & $239.071$  & $180.603$  & $248.437$  & $159.579$  & $\mathbf{90.392}$  & $104.447$  \\ 
		~ & PSNR & $20.606$  & $26.070$  & $28.343$  & $26.165$  & $26.011$  & $26.176$  & $25.960$  & $26.223$  & $26.049$  & $24.510$  & $25.666$  & $24.341$  & $26.268$  & $\mathbf{28.670}$  & $27.984$  \\ 
		~ & SSIM  & $0.508$   & $0.759$   & $0.843$   & $0.747$   & $0.737$   & $0.745$   & $0.739$   & $0.756$   & $0.748$   & $0.715$   & $0.733$   & $0.708$   & $0.750$   & $\mathbf{0.858}$   & $0.840$   \\ \midrule
		
		\multicolumn{1}{c}{\multirow{3}{*}{GOPR0384\_11\_03}} & MSE & $583.758$  & $144.908$  & $84.206$  & $147.814$  & $153.157$  & $148.478$  & $153.906$  & $143.720$  & $147.345$  & $184.127$  & $163.275$  & $170.041$  & $144.334$  & $\mathbf{72.965}$  & $81.592$  \\ 
		~ & PSNR & $20.477$  & $26.532$  & $28.970$  & $26.485$  & $26.329$  & $26.515$  & $26.309$  & $26.688$  & $26.521$  & $25.616$  & $26.050$  & $25.986$  & $26.662$  & $\mathbf{29.641}$  & $29.114$  \\ 
		~ & SSIM  & $0.366$   & $0.636$   & $0.752$   & $0.620$   & $0.609$   & $0.618$   & $0.612$   & $0.636$   & $0.625$   & $0.638$   & $0.602$   & $0.667$   & $0.629$   & $\mathbf{0.797}$   & $0.776$   \\ \midrule
		
		\multicolumn{1}{c}{\multirow{3}{*}{GOPR0869\_11\_00}} & MSE & $596.748$  & $162.904$  & $124.323$  & $227.644$  & $232.791$  & $228.820$  & $237.726$  & $224.742$  & $230.848$  & $360.832$  & $261.173$  & $349.944$  & $223.802$  & $\mathbf{113.112}$  & $113.134$  \\ 
		~ & PSNR & $20.381$  & $26.029$  & $27.280$  & $24.710$  & $24.613$  & $24.721$  & $24.518$  & $24.776$  & $24.648$  & $22.748$  & $24.104$  & $22.923$  & $24.826$  & $\mathbf{27.736}$  & $27.722$  \\ 
		~ & SSIM  & $0.385$   & $0.618$   & $0.685$   & $0.549$   & $0.545$   & $0.551$   & $0.546$   & $0.563$   & $0.549$   & $0.480$   & $0.525$   & $0.506$   & $0.558$   & $\mathbf{0.719}$   & $\mathbf{0.719}$   \\ \midrule
		
		\multicolumn{1}{c}{\multirow{3}{*}{GOPR0396\_11\_00}} & MSE & $586.621$  & $214.140$  & $142.820$  & $188.790$  & $195.849$  & $194.451$  & $200.790$  & $190.093$  & $201.145$  & $387.963$  & $221.714$  & $431.951$  & $186.439$  & $\mathbf{134.292}$  & $164.029$  \\ 
		~ & PSNR & $20.454$  & $24.831$  & $26.754$  & $25.516$  & $25.368$  & $25.433$  & $25.246$  & $25.479$  & $25.243$  & $22.612$  & $24.839$  & $22.118$  & $25.605$  & $\mathbf{27.019}$  & $26.054$  \\ 
		~ & SSIM  & $0.508$   & $0.698$   & $0.790$   & $0.720$   & $0.715$   & $0.719$   & $0.714$   & $0.726$   & $0.715$   & $0.626$   & $0.703$   & $0.605$   & $0.724$   & $\mathbf{0.810}$   & $0.768$   \\ \midrule
		
		\multicolumn{1}{c}{\multirow{3}{*}{GOPR0385\_11\_01}} & MSE & $574.512$  & $114.214$  & $80.911$  & $197.213$  & $200.514$  & $195.313$  & $199.630$  & $191.590$  & $195.861$  & $203.037$  & $207.484$  & $183.812$  & $192.879$  & $\mathbf{70.785}$  & $71.109$  \\ 
		~ & PSNR & $20.555$  & $27.572$  & $29.118$  & $25.373$  & $25.280$  & $25.571$  & $25.381$  & $25.646$  & $25.466$  & $25.413$  & $25.130$  & $25.881$  & $25.609$  & $\mathbf{29.744}$  & $29.720$  \\ 
		~ & SSIM  & $0.337$   & $0.671$   & $0.752$   & $0.542$   & $0.538$   & $0.552$   & $0.545$   & $0.562$   & $0.553$   & $0.588$   & $0.529$   & $0.626$   & $0.553$   & $0.795$   & $\mathbf{0.796}$   \\
		\bottomrule[1.5pt]
	\end{tabular}
}
\end{table*}

\begin{figure*}
	\centering
	\includegraphics[width=7in]{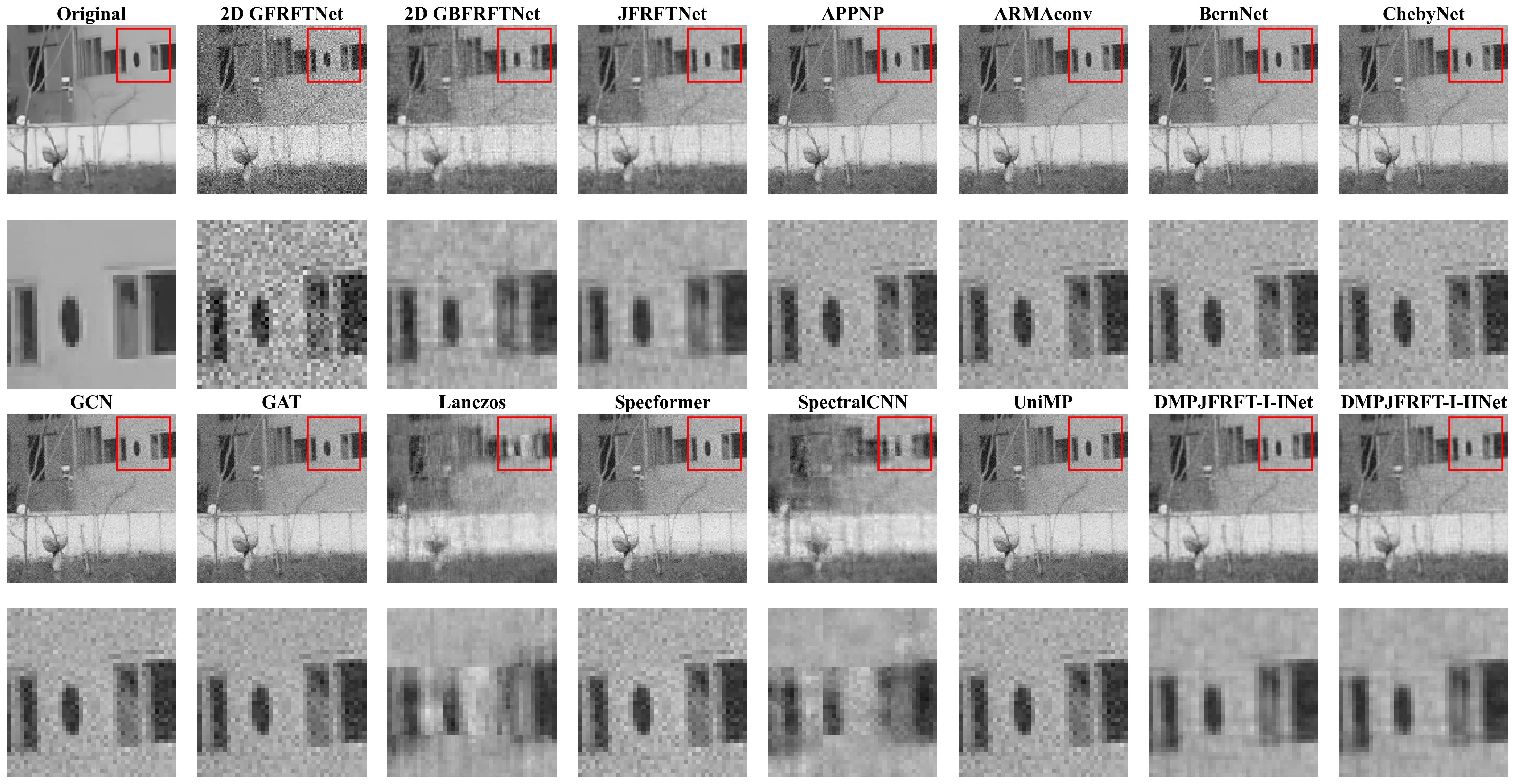}
	\caption{Visual comparison of denoising results for Video GOPR0869\_11\_00.}
	\label{fig:noisy_GOPR0869_11_00}
\end{figure*}

\indent To further demonstrate the visual superiority of the proposed model, Fig.~\ref{fig:noisy_GOPR0869_11_00} shows the denoising results on the $3$rd frame of the $19$th test sample from the GOPR0869\_11\_00 video sequence. As can be seen, the two DMPJFRFTNet variants yield the most restorations, with the least remaining noise and well-preserved structural details. 

\subsection{Debluring}
\begin{figure*}
	\centering
	\includegraphics[width=7in]{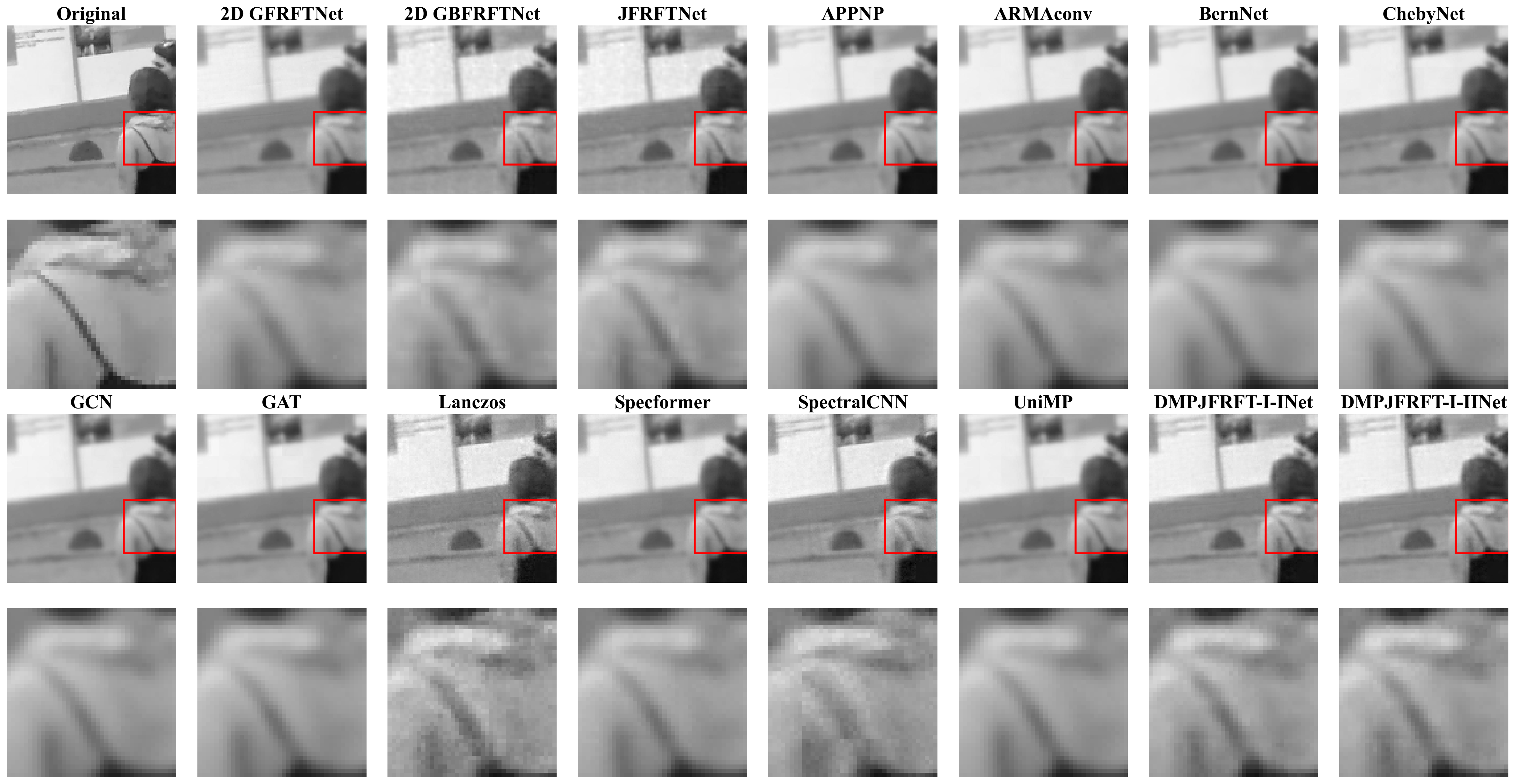}
	\caption{Visual comparison of debluring results for Video GOPR0384\_11\_03.}
	\label{fig:blur_GOPR0384_11_03}
\end{figure*}

\indent In this experiment, the input videos are corrupted by Gaussian blur with kernel size $15$, while all other experimental settings remain the same as those in \emph{2) Video denoising} in Section~\ref{Neural Network Denoising}. Table~\ref{tab:Neural Network deblur viedo} presents the quantitative comparison results of the proposed DMPJFRFTNet and several representative baseline methods in terms of MSE, PSNR, and SSIM, averaged over all video sequences in the test set. The proposed model achieves superior quantitative results, underscoring the benefit of the multiple-parameter formulation in enhancing adaptivity and detail preservation under diverse blur conditions. Fig.~\ref{fig:blur_GOPR0384_11_03} shows the deblurring results on the $4$th frame of the $41$th test sample from the GOPR0384\_11\_03 video sequence. The results indicate that the DMPJFRFTNet generates visually natural restorations with clear structural details and reduced artifacts.

\begin{table*}[htbp]
	\caption{Debluring performance of neural network-based methods on five representative videos from the GoPro dataset.
		\label{tab:Neural Network deblur viedo}}
	\centering
	\resizebox{1.0\linewidth}{!}{
		\begin{tabular}{ccccccccccccccccc}
			\toprule[1.5pt]
			~ & ~ & 2D GFRFTNet & 2D GBFRFTNet & JFRFTNet & APPNP & ARMAconv & BernNet & ChebyNet & GAT & GCN & LanczosNet & Specformer & SpectralCNN & UniMP & DMPJFRFT-I-INet & DMPJFRFT-I-IINet \\ \midrule
			\multicolumn{1}{c}{\multirow{3}{*}{GOPR0384\_11\_02}} &
			MSE & $180.903$  & $147.662$  & $144.495$  & $173.323$  & $179.793$  & $174.213$  & $176.247$  & $172.220$  & $177.756$  & $163.410$  & $177.353$  & $198.772$  & $170.369$  & $134.635$  & $\mathbf{133.289}$  \\ 
			~ & PSNR & $25.590$  & $26.472$  & $26.566$  & $25.795$  & $25.643$  & $25.768$  & $25.744$  & $25.840$  & $25.692$  & $26.256$  & $25.687$  & $25.402$  & $25.889$  & $26.870$  & $\mathbf{26.914}$  \\ 
			~ & SSIM  & $0.824$   & $0.845$   & $0.849$   & $0.838$   & $0.832$   & $0.835$   & $0.836$   & $0.839$   & $0.831$   & $0.820$   & $0.832$   & $0.788$   & $0.840$   & $0.857$   & $\mathbf{0.858}$    \\ \midrule
			
			\multicolumn{1}{c}{\multirow{3}{*}{GOPR0384\_11\_03}} &  MSE & $87.148$  & $74.411$  & $72.512$  & $82.200$  & $84.203$  & $84.529$  & $84.181$  & $82.024$  & $86.288$  & $105.673$  & $85.821$  & $124.151$  & $81.820$  & $67.279$  & $\mathbf{67.086}$  \\ 
			~ & PSNR & $28.839$  & $29.505$  & $29.610$  & $29.075$  & $28.973$  & $28.966$  & $28.976$  & $29.088$  & $28.883$  & $28.100$  & $28.879$  & $27.459$  & $29.113$  & $29.951$  & $\mathbf{29.967}$  \\ 
			~ & SSIM  & $0.870$   & $0.869$   & $0.874$   & $0.879$   & $0.879$   & $0.877$   & $0.876$   & $0.882$   & $0.867$   & $0.806$   & $0.869$   & $0.789$   & $0.881$   & $0.885$   & $\mathbf{0.887}$   \\ \midrule
			
			\multicolumn{1}{c}{\multirow{3}{*}{GOPR0869\_11\_00}} & MSE & $155.789$  & $136.145$  & $120.285$  & $138.934$  & $140.392$  & $144.900$  & $141.549$  & $138.837$  & $144.840$  & $230.827$  & $148.488$  & $306.647$  & $136.746$  & $\mathbf{117.242}$  & $121.866$  \\ 
			~ & PSNR & $26.401$  & $26.971$  & $27.529$  & $26.960$  & $26.909$  & $26.767$  & $26.895$  & $26.977$  & $26.785$  & $24.768$  & $26.623$  & $23.574$  & $27.030$  & $\mathbf{27.643}$  & $27.470$  \\ 
			~ & SSIM  & $0.783$   & $0.776$   & $0.814$   & $0.816$   & $0.815$   & $0.808$   & $0.813$   & $0.817$   & $0.803$   & $0.678$   & $0.798$   & $0.616$   & $0.818$   & $\mathbf{0.822}$   & $0.816$   \\ \midrule
			
			\multicolumn{1}{c}{\multirow{3}{*}{GOPR0396\_11\_00}} & MSE & $286.229$  & $244.930$  & $236.859$  & $265.566$  & $274.670$  & $272.340$  & $267.527$  & $258.628$  & $273.445$  & $307.239$  & $276.074$  & $427.161$  & $259.784$  & $223.677$  & $\mathbf{222.954}$  \\ 
			~ & PSNR & $23.584$  & $24.264$  & $24.401$  & $23.909$  & $23.759$  & $23.798$  & $23.878$  & $24.024$  & $23.796$  & $23.645$  & $23.735$  & $22.160$  & $24.009$  & $24.659$  & $\mathbf{24.679}$  \\ 
			~ & SSIM  & $0.759$   & $0.773$   & $0.782$   & $0.787$   & $0.779$   & $0.779$   & $0.783$   & $0.792$   & $0.776$   & $0.728$   & $0.775$   & $0.651$   & $0.791$   & $0.797$   & $\mathbf{0.798}$   \\ \midrule
			
			\multicolumn{1}{c}{\multirow{3}{*}{GOPR0385\_11\_01}} & MSE & $60.143$  & $48.973$  & $43.879$  & $50.506$  & $51.683$  & $53.672$  & $51.945$  & $50.414$  & $53.276$  & $102.066$  & $52.542$  & $131.189$  & $49.953$  & $\mathbf{41.451}$  & $42.931$  \\ 
			~ & PSNR & $31.036$  & $31.805$  & $32.306$  & $31.801$  & $31.723$  & $31.552$  & $31.679$  & $31.800$  & $31.573$  & $28.767$  & $31.570$  & $27.780$  & $31.864$  & $\mathbf{32.555}$  & $32.403$  \\ 
			~ & SSIM  & $0.888$   & $0.891$   & $0.907$   & $0.908$   & $0.906$   & $0.902$   & $0.906$   & $0.908$   & $0.902$   & $0.793$   & $0.902$   & $0.757$   & $0.909$   & $\mathbf{0.912}$   & $0.910$   \\
			\bottomrule[1.5pt]
		\end{tabular}
	}
\end{table*}

\indent Let $\mathcal{I}_l$ and $\mathcal{O}_l$ represent the input and output feature dimensions of the $l$-th layer, respectively. $\mathcal{H}_l{'}$ and $\mathcal{H}_l{''}$ stand for the hidden dimensions. $h$ denotes the number of MLP units. $K$ indicates the polynomial order for Chebyshev or Bernstein approximations. $A_l$ is the count of attention heads, and $E$ is the number of graph edges. $S$ signifies the number of parallel stacks, while $k$ denotes the propagation depth. $l$ and $s$ correspond to parameters for long and short scales, respectively.

\indent For the $l$-th layer, $\mathcal{I}_l$ and $\mathcal{O}_l$ represent the input and output feature dimensions, respectively, while $\mathcal{H}_l'$ and $\mathcal{H}_l''$ correspond to the intermediate hidden dimensions. Let $h$ denote the number of MLP units. The number of attention heads in this layer is denoted by $A_l$. The polynomial order used in Chebyshev or Bernstein approximations is indicated by $K$, and the total number of edges in the graph is represented by $E$. $S$ denotes the number of parallel stacks employed in the architecture. Regarding propagation-related parameters, $k$ specifies the propagation depth, whereas $\ell$ and $s$ correspond to the long-range and short-range propagation scales, respectively. Table~\ref{tab:method_comparison} presents the key characteristics of different implementations under both gradient descent-based filtering and neural network-based filtering.

\begin{table*}[htbp]
		\caption{Comparison of different filtering methods.
		\label{tab:method_comparison}}
	\centering
	\resizebox{1.0\linewidth}{!}{
		\begin{tabular}{lcccccc}
			\toprule
			\textbf{Method} & \textbf{Parameter Count} & \textbf{Feature Type} & \textbf{Complexity} & \textbf{Prior Information} & \textbf{Eigendecomposition} & \textbf{Dynamic} \\
			\midrule
			\multicolumn{7}{c}{Gradient descent-based filtering} \\
			\midrule
			2D GFRFT & $N + 1$ & Spatio-spatial feature & $\mathcal{O}(N^3)$ & Full & Yes & No \\
			2D GBFRFT & $N + 2$ & Spatio-spatial feature & $\mathcal{O}(N^3)$ & Full & Yes & No \\
			JFRFT & ${N\mathcal{I}} + 2$ & Spatio-temporal feature & $\mathcal{O}({N^3})$ & Full & Yes & No \\
			DMPJFRFT & $2N\mathcal{I} + \mathcal{I}$ & Spatio-temporal feature & $\mathcal{O}({N^3})$ & Full & Yes & Yes \\
			\midrule
			\multicolumn{7}{c}{Neural network-based filtering} \\
			\midrule
			2D GFRFTNet & $N + 1$ & Spatio-spatial feature & $\mathcal{O}(N^3)$ & Partial & Yes & No \\
			2D GBFRFTNet & $N + 2$ & Spatio-spatial feature & $\mathcal{O}(N^3)$ & Partial & Yes & No \\
			JFRFTNet & ${N\mathcal{I}} + 2$ & Spatio-temporal feature & $\mathcal{O}({N^3})$ & Partial & Yes & No \\
			APPNP & $\mathcal{I}_l {\mathcal{O}_l}+\mathcal{O}_l$ & Spatial feature & $\mathcal{O}(N\mathcal{I}\mathcal{O} + kE\mathcal{I})$ & Partial  & No & No \\
			ARMAconv & $S(2\mathcal{I}_l \mathcal{O}_l+(\mathcal{O}_l)^2+\mathcal{O}_l$ & Spatial feature & $\mathcal{O}(Sk(2N\mathcal{I}\mathcal{O}+E\mathcal{I}))$ & Partial & No & No \\
			BernNet & $K+1+\mathcal{I}_l \mathcal{O}_l + \mathcal{O}_l$ & Spatial feature & $\mathcal{O}(N\mathcal{I}\mathcal{O} + KE\mathcal{I})$ & Partial & No & No \\
			ChebyNet & $(K + 1) \mathcal{I}_l \mathcal{O}_l + \mathcal{O}_l$ & Spatial feature & $\mathcal{O}(N\mathcal{I}\mathcal{O} + KE\mathcal{I})$ & Partial  & No & No \\
			GAT & $A_l (\mathcal{I}_l {\mathcal{O}_l} + 2\mathcal{O}_l) + A_l \mathcal{O}_l$ & Spatial feature & $\mathcal{O}(HN\mathcal{I}\mathcal{O} + HE\mathcal{I})$ & Partial & No & No \\
			GCN & $\mathcal{I}_l \mathcal{O}_l + \mathcal{O}_l$ & Spatial feature & $\mathcal{O}(N\mathcal{I}\mathcal{O} + E\mathcal{I})$ & Partial & No  & No \\
			LanczosNet & $(2\ell+2)h+\ell+\mathcal{O}_l+h^2+(\ell+s)(\mathcal{O}_l)^2$ & Spatial feature & $\mathcal{O}(N^3)$ & Partial & Yes & No \\
			Specformer & $3(\mathcal{O}_l)^2+\mathcal{O}_l( {\mathcal{I}_l}+\mathcal{H}'_l+2\mathcal{H}_l''+5)+\mathcal{H}_l'+\mathcal{H}_l''$ & Spatial feature & $\mathcal{O}({N^3})$ & Partial & Yes & No \\
			SpectralCNN & $\mathcal{I}_l \mathcal{O}_l+\mathcal{O}_l\mathcal{H}_l'N+\mathcal{H}_l'\mathcal{H}_l''$ & Spatial feature & $\mathcal{O}(N^3)$ & Partial  & Yes & No \\
			UniMP & $(4\mathcal{I}_l \mathcal{O}_l+{\mathcal{O}_l}){H}_l$ & Spatial feature & $\mathcal{O}({N})$ & Partial & No & No \\
			DMPJFRFTNet & $2N\mathcal{I} + \mathcal{I}$ & Spatio-temporal feature & $\mathcal{O}({N^3})$ & Partial & Yes & Yes \\
			\bottomrule
		\end{tabular}
	}
\end{table*}

\section{Conclusion}   \label{sec6}
\indent In this paper, we introduced the DMPJFRFT, a novel framework for dynamic graph signal processing. Instead of using a single fractional order in each domain like JFRFT, the proposed DMPJFRFT leverages multiple-parameter mechanism to perform graph structure modeling in the spectral domain, thereby capturing the evolving topology and temporal dynamics of graph signals. To validate the effectiveness of the proposed framework, we further developed two dynamic filtering strategies, namely a gradient descent-based filtering approach and a neural network implementation referred to as DMPJFRFTNet. Experimental evaluations on both real-world dynamic graph data and video datasets demonstrate that the DMPJFRFT achieves effective performance in denoising and deblurring tasks.



\bibliography{mybib}
\bibliographystyle{IEEEtran}

\begin{IEEEbiographynophoto}{Manjun~Cui}
	received the B.S. degree in Mathematics and Applied Mathematics from Yancheng Teachers University, Yancheng, Jiangsu, China, in 2022. She is currently pursuing the Ph.D. degree in mathematics with the School of Mathematics and Statistics, Nanjing University of Information Science and Technology, Nanjing, Jiangsu, China.
\end{IEEEbiographynophoto}

\begin{IEEEbiographynophoto}{Ziqi~Yan}
	received the B.S. degree in Mathematics and Applied Mathematics from Yancheng Teachers University, Yancheng, Jiangsu, China, in 2023. He is currently pursuing the M.S. degree in mathematics with the School of Mathematics and Statistics, Nanjing University of Information Science and Technology, Nanjing, Jiangsu, China.
\end{IEEEbiographynophoto}

\begin{IEEEbiographynophoto}{Yangfan~He}
	received the Ph.D. degree in space physics from Wuhan University, Wuhan, China, in 2022. She is currently a Lecturer with the School of Information and Communication Engineering, Nanjing Institute of Technology, Nanjing, Jiangsu, China. Her current research interests include signal processing, plasma waves, and magnetosphere-ionosphere coupling.
\end{IEEEbiographynophoto}

\begin{IEEEbiographynophoto}{Zhichao~Zhang}
	(Member, IEEE) received the B.S. degree in mathematics and applied mathematics from Gannan Normal University, Ganzhou, Jiangxi, China, in 2012, and the Ph.D. degree in mathematics of uncertainty processing from Sichuan University, Chengdu, Sichuan, China, in 2018. From September 2017 to September 2018, he was a Visiting Student Researcher with the Department of Electrical and Computer Engineering, Tandon School of Engineering, New York University, Brooklyn, NY, USA, where he was awarded a grant from the China Scholarship Council. Since 2019, he has been with the School of Mathematics and Statistics, Nanjing University of Information Science and Technology, Nanjing, Jiangsu, China, where he is currently a Full Professor and a Ph.D Supervisor. From January 2021 to January 2023, he was a Macau Young Scholars Post-Doctoral Fellow of information and communication engineering with the School of Computer Science and Engineering, Macau University of Science and Technology, Macau, SAR, China. He has published more than 60 journal articles in IEEE TRANSACTIONS ON INFORMATION THEORY, IEEE TRANSACTIONS ON SIGNAL PROCESSING, IEEE SIGNAL PROCESSING LETTERS, IEEE COMMUNICATIONS LETTERS, Signal Processing, and Journal of Fourier Analysis and Applications. His current research interests include mathematical theories, methods, and applications in signal and information processing, including fundamental theories, such as Fourier analysis, functional analysis and harmonic analysis, applied theories, such as signal representation, sampling, reconstruction, filter, separation, detection and estimation, and engineering technologies, such as satellite communications, radar detection, and electronic countermeasures. He was a member of the International Association of Engineers, the China Society for Industrial and Applied Mathematics, the Chinese Institute of Electronics, and the Beijing Society for Interdisciplinary Science. He was the Vice President of the Jiangsu Society for Computational Mathematics and the Director of the Jiangsu Society for Industrial and Applied Mathematics. He was listed among world's top 2\% scientists recognized by Stanford University in 2021 and 2022.
\end{IEEEbiographynophoto}

\end{document}